\documentclass[nofootinbib,eqsecnum,tightenlines,superscriptaddress,10pt]{revtex4}

\usepackage{graphicx}
\usepackage{wrapfig,lipsum}
\usepackage{amsmath,amssymb,amsfonts,amsthm,stmaryrd,mathtools,bm}
\usepackage{mathrsfs}
\usepackage{color}
\usepackage{multirow,bigdelim}
\usepackage{hyperref}
\usepackage{dsfont}
\usepackage{booktabs}
\allowdisplaybreaks[0]

\usepackage{pstricks}
\usepackage{tikz}
\usepackage{ulem}

\def\be#1\ee{\begin{align}#1\end{align}}

\def\ba{\begin{eqnarray}}
\def\ea{\end{eqnarray}}
\def\nn{\nonumber}
\def\q{\quad}

\begin{document}

\title{Towards effective actions for the continuum limit of spin foams}

\author{Johanna N.~Borissova}
\affiliation{Perimeter Institute, 31 Caroline Street North, Waterloo, ON, N2L 2Y5, CAN}
\affiliation{Department of Physics and  Astronomy, University of Waterloo, 200 University Avenue West, Waterloo, ON, N2L 3G1, Canada}
\author{Bianca Dittrich}
\affiliation{Perimeter Institute, 31 Caroline Street North, Waterloo, ON, N2L 2Y5, CAN}

\begin{abstract}

Spin foams arise from a quantization of classical gravity expressed via the Plebanski action. Key open questions related to the continuum limit of spin foams are whether general relativity is reproduced and what type of corrections could emerge. As a central component for spin foam dynamics, recent results on the continuum limit of the Area Regge action suggest a close relation with actions for area metrics instead of a length metric. Inspired by these results, within the framework of modified Plebanski theory we construct a family of candidate actions for area metrics.   These actions are expected to describe the continuum limit of spin foams and provide a starting point to explore phenomenological aspects of the large-scale dynamics of spin foams. More generally, they set the stage for exploring consequences of an enlargement of the configuration space for gravity from length to area metrics. The actions we construct lead to an effective action for the length metric, describing a non-local and ghost-free version of Einstein-Weyl gravity.

\end{abstract}

\maketitle


\section{Introduction}\label{Intro}

Spin foams \cite{PerezReview} constitute a path integral approach to quantum gravity, based on the Plebanski action \cite{Plebanski,Chiral,NonCh,Reisi}.  Instead of a length metric as in the Einstein-Hilbert action, the basic fields in the Plebanski action are given by an $\text{so}(4)$-valued\footnote{We will be concerned with four-dimensional space-times, and for simplicity consider Euclidean signature.} two-form $B$, as well as a conjugated $\text{SO}(4)$-connection. 

This has allowed for  background independent and non-perturbative constructions of the path integral \cite{PerezReview}. One of the reasons for this success is that the Plebanski action is closely related to topological $BF$-theories \cite{Baez}, which can be quantized exactly, that is preserving their topological nature and respecting their many symmetries.  

 But the use of the Plebanski formulation has also made certain questions more difficult to answer.   This includes the continuum limit of the theory and its interpretation in terms of geometric quantities.  The main aim of this work is to provide well motivated candidate actions for the description of the continuum limit of the theory. 
 
A central assumption in this work is that the relevant geometric quantitiy describing the continuum limit of spin foams is given by an area metric and not a length metric. Similar to the length metric measuring the length of tangent vectors and angles between tangent vectors, the area metric measures the areas of parallelograms spanned by two tangent vectors, and dihedral angles between such parallelograms.  In four space-time dimensions the cyclic\footnote{The cyclicity condition will be explained and motived in Section \ref{SecArea}.} area metric has 20 independent components, that is double as many independent components as the length metric. We have  thus an extended geometric configuration space.  

Area metrics appear not only in the spin foam approach to quantum gravity, but might also arise as a coarse grained geometric quantity from string theory \cite{Schuller1}, or for approaches where geometry is reconstructed from entanglement \cite{RyuTakayanagi}. Area metric space-times and actions have been constructed in \cite{Schuller2}. We will here take a different approach than in \cite{Schuller2}, which will result in a different class\footnote{The action constructed in \cite{Schuller2} is homogeneous in second derivatives, whereas our action will feature mass terms for the degrees of freedom that are not induced from a length metric. These mass terms are essential for recovering to leading order general relativity in the effective action for the length metric.} of area metric actions.

There are a number of good reasons to consider area metrics for the continuum limit of spin foams, which we will review in more detail in Section \ref{SecMo}. One key reason is rooted in the simplicity constraints, which are a central mechanism in the Plebanski formalism. Their implementation turns the topological $BF$-theory into a gravitational theory and reduces the $B$-fields to tetrads, whose gauge reduction gives the length metric.  But quantization leaves the simplicity constraints  partially anomalous and a part of the simplicity constraints are still allowed to fluctuate. This leads to an extension of the configuration space for the path integral: one is not only summing over length metric degrees of freedom, but also over degrees of freedom describing the fluctuating part of the simplicity constraints. 

Another key reason for suspecting that the space of area metrics constitutes the relevant extension are the findings of the recents works \cite{AR1,AR2}, which examine the continuum limit of Area Regge calculus. The Area Regge action appears in the semi-classical analysis of spin foam amplitudes, but a description of its dynamics in terms of continuum fields remained elusive for a long time. \cite{AR1,AR2} found that Area Regge calculus expanded on a hyper-cubic lattice gives at leading order in the lattice constant the dynamics of (linearized) general relativity. Moreover, \cite{AR2} identified the leading order correction as resulting from having an area metric instead of a length metric as underlying field.  A crucial point in the mechanism of obtaining general relativity at leading order from the Area Regge action, is that this action produces mass terms for the degrees of freedom in the area metric that cannot be induced from a length metric.
 Integrating out these additional degrees of freedom in the area metric, one obtains an effective action for the length metric. This effective action contains the Einstein-Hilbert term, and at second order in the lattice constant, the square of the Weyl curvature.

The works \cite{AR1,AR2} examined the continuum limit of Area Regge calculus, which arises in the semi-classical limit of spin foams, which are, in turn, quantizations of the Plebanski action. A natural question --- and central to this work --- is therefore, whether one can derive more directly effective actions as the one obtained from Area Regge calculus, from the Plebanski action.  As the latter is the starting point of spin foam quantization, it is reasonable to assume that the continuum limit of spin foams is described by the Plebanski action or a modification thereof.

The Plebanski action itself, with all its simplicity constraints imposed sharply, leads\footnote{More precisely, the simplicity constraints admit several sectors of solutions and one of these sectors leads to general relativity.} to general relativity. But as we have discussed, the spin foam quantization procedure does not allow for the sharp implementation of all the simplicity constraints. We are therefore looking for a way to impose one part of these constraints sharply and another part only weakly.

Indeed, similar modifications of the Plebanski action have been proposed by Krasnov \cite{Krasnov1}, and are known as modified Plebanski actions.  In these modified Plebanski actions one replaces the simplicity constraints by a potential which suppresses these constraints but still allows them to fluctuate. The potential can be chosen such that a certain subset of the constraints is still imposed sharply, whereas the remainder of the constraints are only suppressed e.g. via  mass terms. 

It was found \cite{Krasnov1,FreidelMod,Krasnov2} that for chiral\footnote{Chirality refers to the decomposition $\text{so}(4)=\text{su}(2) \oplus \text{su}(2)$. Using a chiral version can be understood as imposing part of the simplicity constraints, reducing the $\text{so}(4)$-valued fields to the left or right $\text{su}(2)$-part.} versions of the Plebanski actions, where one works with $\text{su}(2)$-valued $B$-fields instead of $\text{so}(4)$-valued ones, a general choice of potential does surprisingly still lead to a `deformation' of general relativity, in the sense that one still has just the spin-2 graviton as propagating degree of freedom\footnote{These theories are related to previously constructed deformations of general relativity \cite{Beng}.}.  There are 5 additional fields, which however do (after a non-local field redefinition of the length metric field) only come with a mass term and not with a kinetic term. These are therefore referred to as auxiliary fields.

We will however work with the non-chiral version of the Plebanski action, as done in spin foams. (Another reason is that the $B$-fields in the chiral version do not feature sufficiently many degrees of freedom to lead to an area metric with 20 independent degrees of freedom.)  A general modification for the non-chiral version does however lead to a massless and a massive graviton \cite{AlexandrovK,Spez,SpezB} as propagating degrees of freedom (again after a non-local redefinition of the length metric field). This corresponds to a bi-metric theory \cite{Spez} and hence 20 degrees of freedom. There are in addition 10 non-propagating  auxiliary fields, which are massive.

The above lead over-all to 30 degrees of freedom, which are clearly too many to fit into the 20 degrees of freedom of the area metric. But a general choice of the potential means that all the simplicity constraints are imposed only weakly, that is via a potential. We have discussed above, that spin foams do impose a (commuting) subset of the simplicity constraints sharply, whereas the remainder is imposed weakly. 

A natural\footnote{Another choice, which leads to 20 degrees of freedom is to impose that the auxiliary fields vanish, leaving the two metrics. But although one of the gravitons is massive, this theory does not feature 10 massive degrees of freedom.} choice, which leads to 20 degrees of freedom, is to impose sharply that the two length metrics in the bi-metric description, actually coincide. We therefore will get one length metric with 10 massless degrees of freedom, and 10 auxiliary fields with mass, as in Area Regge calculus.  This split between sharply and weakly imposed simplicity constraints will be the central assumption in our paper.

We will $(a)$ show that these 20 degrees of freedom can be indeed packaged into an area metric and that $(b)$ the corresponding modified Plebanski action leads to the same effective action for the length metric as Area Regge calculus. Curiously, we will however find that the actions for the area metrics as defined from the modified Plebanski approach and the Area Regge action are structurally very similar but do differ in a sign for the coupling between the length metric and remaining degrees of freedom. 

Apart from the ideas outlined above, the work here benefits from two further inputs. Firstly, the concept of area metrics has appeared previously in an analysis of the Plebanski action by Reisenberger in \cite{Reisi}. There, the $\text{so}(4)$-valued $B$-fields are split into  left-handed and right handed $\text{su}(2)$-valued $B$-fields respectively.  A left-handed and right-handed area metric is defined from these two sectors. A key result of \cite{Reisi} is, that the simplicity constraints are (locally) equivalent to demanding that these two area metrics are equal. 

Note however, that the left- and right- handed area metrics are not well suited for a variable transformation: the left- and right-handed $B$-fields contain each 15 independent $\text{su}(2)$ gauge invariant degrees of freedom, whereas the area metrics have each 20 independent components.  

But imposing that the left-handed area metric is equal to the right-handed one, leads to 20 conditions, reducing the 30 gauge invariant degrees of freedom of the $B$-fields to the 10 degrees of freedom of a length metric.  The above choice of split of the set of simplicity constraints, means that we impose only that the length metric parts, defined from the left- and right-handed area metrics respectively, have to coincide. This leads to 10 conditions, leaving us with 20 degrees of freedom. These can be described by an area metric defined from the sum of the left-handed and right-handed one. 

In other words, we will use a modified Plebanski action, in which a part of the simplicity constraints is imposed sharply. This part demands that the left-handed length geometry is equal to the right-handed length geometry. The remaining degrees of freedom can be organized into one area metric. One can in this way obtain an action for area metrics.

The second input we will benefit from, is a parametrization of the $B$-fields, introduced by Freidel in \cite{FreidelMod}, and applied by Speziale in \cite{Spez} to the non-chiral case. This parametrization applies to the left-handed and right-handed $B$-fields respectively, and splits each of these --- on the gauge invariant level --- into a length metric and five further auxiliary fields. The conformal class of the length metric coincides with the Urbantke metric \cite{Urbantke}, that can be defined from the $B$-fields.  

This allows us to define a similar split of the area metric, which will be parametrized by a length metric and 10 additional fields. This provides a natural procedure to extract a length metric from an area metric\footnote{See \cite{Schuller1} for a discussion of area metrics in terms of several length metrics, and \cite{Schuller2} for an alternative procedure to extract an effective length metric from the area metric. This procedure seems however to involve derivatives of the area metric.}, although an analogue of the Urbantke formula for area metrics might be available only perturbatively.    With such a split at hand it is straightforward to design area metric actions, which to leading order give general relativity, and to fine-tune the correction terms.

~\\
 This paper is structured as follows: In Section \ref{SecMo} we will review a number of reasons, why spin foams, and more generally loop quantum gravity, lead to an extended configuration space, as compared with gravity based on a length metric.  We will in particular motivate, why this extension can be described by an area metric. Section \ref{SecArea} introduces (cyclic) area metrics and their algebraic symmetries. We will in particular explain, how the assumption that we can only measure areas of all possible parallelograms in tangent space (but not independently angles), leads to cyclic area metrics. We then provide in Section \ref{SecMPleb} a short review of the Plebanski action and its modification.
 
 A key section is Section \ref{Bparam} in which we firstly review how the $B$-fields can be parametrized via two length metrics (or tetrads) and additionally 10 auxiliary fields. This also allows a convenient split of the simplicity constraints into two sets. We will impose sharply the subset of the simplicity constraints, which demands that the two tetrads have to coincide. We then construct an area metric from the remaining degrees of freedom. We finally consider a linearization, which allows us to explicitly invert the area metric fluctuations for the length metric fluctuations and the auxiliary fields.
 
 We then discuss in Section \ref{SecAct} the (linearized) actions as defined by the modified Plebanski framework. The linearization helps to uncover its dynamical content. The version of the modified Plebanski actions we define, does not allow anymore a re-definition of the fields so that only the length metric propagates, as done for the modifications of the chiral Plebanski action. However, we will see that the part quadratic in the length metric reproduces the linearized Einstein-Hilbert action and that the auxiliary fields will be coupled to the Weyl-curvature of the length metric and equipped with a mass term.  We can integrate out the auxiliary fields, which leads to an effective action for the length metric.  This will lead to a correction given by the square of the Weyl-curvature.
 
We review in Section \ref{SecAR} the linearized Area Regge action. Comparing this action (as far as it has been determined) with the one derived from the modified Plebanski approach, we will see that it leads to the same effective action for the length metric. The two actions do however come with a different sign for the coupling between the auxiliary field and the length metric.  We thus see that there could be different choices for a suitable area metric action. 

We will close with a discussion and outlook in Section \ref{Disc}.

~\\
Let us mention our main conventions here.  We will be concerned with four-dimensional space-times with Euclidean signature. We will use lower-case greek letters for space-time indices with range $\mu=0,\ldots, 3$. The expressions $\epsilon^{\mu\nu \rho\sigma}$ and $\epsilon_{\mu\nu \rho\sigma}$ are both defined by the Levi-Civita-symbol, and thus tensor densities of weight $+1$, respectively $-1$. To avoid confusion, we do {\it not} use the space-time metric to raise or lower indices for these tensor densities.

We will use anti-symmetric index pairs $(IJ)$ to label a basis of the $\text{so}(4)$-algebra.  Here $I,J=0,\ldots 3$. This algebra admits two bilinear forms given by $\delta_{IJKL}:=\tfrac{1}{2} (\delta_{IK}\delta_{JL}-\delta_{IL}\delta_{JK})$ and $\tfrac{1}{2}\epsilon_{IJKL}=\tfrac{1}{2}\epsilon_{KLIJ}$.  Here we can raise and lower indices with an internal metric $\delta_{IJ}$.

We will use lower case latin letters  $i,j,k$ for  indices  that run from $1,\ldots 3$ and label a basis of the $\text{su}(2)$-algebra.

\section{Motivation for area metrics from spin foams}\label{SecMo}

In the introduction we shortly mentioned two reasons why spin foams lead to an enlarged configuration space, and why we believe that in the continuum limit this enlargement can be captured by using area metrics. Here we will provide a list of more detailed arguments:

\begin{itemize}
\item One central mechanism in the Plebanski formuation of general relativity is the imposition of the so-called simplicity constraints. Imposing these constraints the $B$-fields can be parametrized by tetrads, which modulo $\text{SO}(4)$ gauge symmetry and orientation sectors, carry the same information as the length metric. Whereas the (primary) simplicity constraints are classically first class, the spin foam quantization (with a so-called Holst term) of the $B$-fields in terms of combinations of $\text{so}(4)$-generators leads to an algebra that is partially second class \cite{PerezReview,DittrichRyan1}. That is, we encounter an anomaly which is parametrized by the Barbero-Immirzi  parameter \cite{BI}.  The EPRL/FK spin foam models \cite{EPRL-FK}, as well as the effective spin foam models \cite{EffSF1,EffSF2,EffSF3}  impose the anomalous part of the simplicity constraints weakly:  the corresponding degrees of freedom are still allowed to fluctuate but are suppressed.  We therefore have to deal with an enlargement of the configuration space. The degree to which these degrees of freedom are suppressed is controlled by the Barbero-Immirzi  parameter, which can be understood as anomaly parameter \cite{DittrichRyan3,EffSF2}.

\item A more direct way to notice this enlargement of the configuration space is as follows. This reasoning, which motivated the construction of the effective spin foam models \cite{EffSF1,EffSF2,EffSF3}, applies to the Barrett-Crane models \cite{BC,BO} as well as the  EPRL/FK models \cite{EPRL-FK}.  Spin foam quantization proceeds by triangulating the space-time manifold, and assigning quantum numbers to the triangulation.  A central result of loop quantum gravity \cite{LQG}, which underlies spin foams, is that $(a)$ the areas are independent variables and $(b)$ that the spectra of area operators are discrete \cite{RovelliSmolin}, and that the spectra are asymptotically equi-distant. Typically, there are far more triangles than edges in a four-dimensional triangulation, and thus far more area variables than length variables. Imposing that the area-variables arise from length variables, can therefore be seen as part of the simplicity constraints \cite{AreaAngle,DittrichRyan1}.  However, this part of the constraints constitutes diophantine equations in the area quantum numbers \cite{EffSF1}, which admit too few solutions for a sensible semi-classical limit \cite{EffSF1}.  In fact, having areas as independent variables, gives as a strong reason to introduce an area metric  for the description of the continuum limit.

Similarly, in approaches where geometries are determined from entanglement, e.g. in tensor network states \cite{Swingle2} or bit threads \cite{Headrick}, areas might arise as relevant geometric quantities for the effective description. 

\item  Spin foam related configuration spaces associated to triangulations have been analyzed on the  covariant \cite{AreaAngle,BonzomAreaAngle} and canonical \cite{DittrichRyan1,DittrichRyan2,Twisted} level, and also suggest an enlargement of the geometric configuration space.  The work \cite{AreaAngle} constructs the so-called Area-Angle Regge action and proposes that it describes the classical limit of spin foams. The variables of this action are  areas $A_t$ of triangles and dihedral angles $\phi_{\tau,(t,t')}$ between pairs $(t,t')$ of triangles belonging to the same tetrahedron $\tau$.  To a given four-simplex one can associate 10 area variables and 10 independent dihedral angles (we impose the closure constraints for the tetrahedra). These can be matched with the degrees of freedom of an area metric, associated to that four-simplex \cite{JoseTA}.  

The dihedral angles can be seen as auxiliary variables, which allow to express the condition that the areas result from a consistent length configuration, via constraints that can be localized to the four-simplices \cite{AreaAngle}. These constraints are called gluing or shape-matching constraints and can be understood as follows:  given the 10 areas of a flat (and shape-matched) 4-simplex $\sigma$, one can find  its 10 lengths and thus also determine the values of the three-dimensional dihedral angles $\Phi^{\sigma}_\tau(A)$ associated to the tetrahedra $\tau$ of this 4-simplex $\sigma$. 
A given tetrahedron $\tau$ in the bulk of the triangulation is shared by two 4-simplices. 
Let these be $\sigma$ and $\sigma'$.  The shape-matching constraints \cite{EffSF1,EffSF2} ensure that the dihedral angles in the tetrahedron $\tau$ are consistently defined:  
Firstly, the values of the dihedral angles computed from the areas of the simplex $\sigma$ coincide with the values as computed from the simplex  $\sigma'$, that is 
$\Phi^{\sigma}_\tau(A)=\Phi^{\sigma'}_\tau(A)$. Secondly, the dihedral angle variable is equal to these values $\phi_\tau=\Phi^{\sigma}_\tau(A)=\Phi^{\sigma'}_\tau(A)$.

\item
The canonical analysis \cite{DittrichRyan1,DittrichRyan2} revealed that these conditions are however second class, that is, they cannot be imposed sharply in the quantum theory. (This is made explicit in the effective spin foam construction \cite{EffSF1,EffSF2,EffSF3}.) Moreover, \cite{DittrichRyan1} showed that the phase space associated to this Area-Angle Regge action, does indeed match the phase space of loop quantum gravity, but constitutes a proper enlargement of the (length) Regge phase space. The latter arises from a canonical analysis \cite{Hoehn} of Length Regge calculus, in which the basic configuration variables are lengths associated to the edges of the triangulation. Regge calculus \cite{Regge} with length variables reproduces general relativity in the continuum limit, see e.g. \cite{RocekWilliams}.

\item The Area Regge action \cite{BarrettWilliams, ADH1} arises in the semi-classical limit of spin foams \cite{SFLimit}, and uses only areas as basic variables.  It plays also a central role in the effective spin foam models \cite{EffSF1,EffSF2,EffSF3}.
The Area Regge action can be understood as resulting from the Area-Angle action, if we (a) relax the shape matching constraints by introducing two copies  $\phi^{\sigma}_\tau,\phi^{\sigma'}_\tau $ of the dihedral  angles $\phi_\tau$ --- one for each four-simplex sharing $\tau$. We then (b) replace the shape matching conditions $\phi_\tau=\Phi^{\sigma}_\tau(A)=\Phi^{\sigma'}_\tau(A)$ by the weaker set $\phi^{\sigma}_\tau=\Phi^{\sigma}_\tau(A),\, \phi^{\sigma'}_\tau=\Phi^{\sigma'}_\tau(A)$ and (c) eliminate in this way the dihedral angle variables. 

Although the Area Regge action is central to the semi-classical limit of spin foams, its dynamical content remained an open issue for a long time \cite{Wainwright,ADH1}.  The discrete equations of motion seem\footnote{This impression results if one misinterprets the deficit angle $\epsilon_t(A)=2\pi-\sum_{\sigma \supset t} \theta^\sigma_t(A)$, with $\theta^\sigma_t$ the four-dimensional dihedral angles between tetrahedra, as measuring curvature. But it actually measures a combination of curvature and shape-mismatching \cite{ADH1}, which is difficult to disentangle at the non-linear level.} to suggest \cite{BarrettWilliams} that curvature vanishes. This issue is closely related to the so-called flatness problem of spin foams \cite{flatness}. Indeed, in the Plebanski formalism one obtains general relativity from a topological $BF$-theory (whose equations of motion demand vanishing curvature as defined by the connection) by imposing the simplicity constraints. Imposing part of these constraints only weakly, could this affect the resulting equations of motion, so that these demand flatness?

Recent work \cite{AR1,AR2} revealed however, that the continuum limit of (linearized) Area Regge action, on a regular hyper-cubical lattice, leads to leading order in the lattice constant, to the discretized (linearized) Einstein-Hilbert action (which can be also obtained from the Length Regge action \cite{RocekWilliams}).  The leading order correction to this Einstein-Hilbert action can moreover be understood to result from an underlying area metric \cite{AR2}. Integrating out the degrees of freedom in the area metric which are not induced from the length metric, one obtains an effective action for the length metric, given by the Einstein-Hilbert term and the square of the Weyl curvature \cite{AR2}. We will provide a short review of this result in Section \ref{SecAR}.

These results indicate that the continuum limit of spin foams might provide a surprising resolution for the flatness problem: the degrees of freedom, that we want to suppress by the simplicity constraints are suppressed by a dynamical mechanism, which kicks in when taking the continuum limit\footnote{Here the scaling of certain blocks of the lattice Hessian in the lattice momenta is essential:  the block diagonal in the length metric degrees of freedom scales with the square of the momenta and is massless. The block diagonal in the additional area metric degrees of freedom features mass terms.}.  We furthermore see that the leading order correction in this continuum limit can be described in terms of an area metric. This area metric is constructed from the hyper-cubical building blocks, which each contain 48 four-simplices.  This area metric has thus a more macroscopic origin, than the one constructed from the single four-simplices, which we discussed above.

\end{itemize}

\section{Area metrics}\label{SecArea}

In the same way as we use the length metric to measure the lengths of vectors we can introduce an area metric to measure the areas of parallelograms, described by an outer product of two vectors $(v\wedge w)^{\mu\nu}=-(v\wedge w)^{\nu\mu}$. The area metric $A_{\mu\nu\rho\sigma}$ should therefore have the following symmetries
\ba\label{SymA}
A_{\mu\nu\rho\sigma}=-A_{\nu\sigma\rho\sigma}=-A_{\mu\nu\sigma\rho}=A_{\rho\sigma\mu\nu} \q .
\ea
This gives (in four dimension) 21 independent components. We will however also demand cyclicity, that is
\ba\label{AM22}
A_{\mu\nu\rho\sigma}+A_{\mu\rho\sigma\nu}+A_{\mu\sigma\nu\rho}=0      \q ,
\ea
which, assuming the other symmetries hold, is equivalent to one condition given by
\ba\label{Cycl3.3}
A_{\mu\nu\rho\sigma} \epsilon^{\mu\nu\rho\sigma}=0 \q .
\ea
The cyclic area metric tensor has thus only 20 independent components, and features the same algebraic symmetries as the Riemann tensor.

The cyclicity condition can be motivated as follows\footnote{The cyclicity condition has been motivated in \cite{Schuller1}, using the demand that the area metrics provide an irreducible representation for $\text{SL}(4)$.  We present here a more intuitive argument.}: The length metric can be reconstructed from measuring the length of all vectors only. That is, it is not necessary to measure angles. More precisely, given the basis $\{\delta_\mu^\nu\}_\nu$ in tangent space, we determine the length of its vectors and of the sum of two basis vectors
\ba\label{AM24}
l_{\rho}:=&g_{\mu\nu}\delta^\mu_\rho \delta^\nu_\rho &= g_{\rho\rho}             \nn\\
l_{\rho\sigma}:=&g_{\mu\nu}(\delta^\mu_\rho+\delta^\mu_\sigma)(\delta^\nu_\rho+\delta^\nu_\sigma)&= g_{\rho\rho} +g_{\sigma\sigma} + 2g_{\rho\sigma} 
\ea
Clearly, we can reconstruct all metric components $g_{\rho\sigma}$ from the $l_\rho$ and $l_{\rho\sigma}$.

We demand similarly  that the area metric can be determined by measuring the areas of parallelograms.  Considering those\footnote{Other cases, e.g. considering the sum of three basis vectors, give combinations of the ones considered below.} that are spanned by the basis vectors and by sums of two basis vectors, we have
\ba\label{AM25}
p_{\rho\sigma}:=&A_{\mu\nu\mu'\nu'} \,\,\delta^\mu_\rho   \delta^\nu_\sigma \delta^{\mu'}_\rho \delta^{\nu'}_\sigma &= A_{\rho\sigma\rho\sigma} \nn\\
p_{\rho\sigma\kappa}:=&A_{\mu\nu\mu'\nu'}\,\,\delta^\mu_\rho   (\delta^\nu_\sigma  +\delta^\nu_\kappa ) \delta^{\mu'}_\rho (\delta^{\nu'}_\sigma+\delta^{\nu'}_\kappa) &= A_{\rho\sigma\rho\sigma} +A_{\rho\kappa\rho\kappa} + 2A_{\rho\sigma\rho\kappa}\nn\\
p_{\rho\lambda\sigma\kappa}:=&A_{\mu\nu\mu'\nu'}(\delta^\mu_\rho  +\delta^\mu_\lambda) \,\,(\delta^\nu_\sigma  +\delta^\nu_\kappa ) (\delta^{\mu'}_\rho +\delta^{\mu'}_\lambda)(\delta^{\nu'}_\sigma+\delta^{\nu'}_\kappa) &= A_{\rho\sigma\rho\sigma} +A_{\rho\kappa\rho\kappa} + A_{\lambda\sigma\lambda\sigma}  + A_{\lambda\kappa\lambda\kappa} + \nn\\
&&\q\; 2A_{\rho\sigma\rho\kappa}+2A_{\lambda\sigma\lambda\kappa}+2A_{\rho\sigma\lambda\sigma}+2A_{\rho\kappa\lambda\kappa}+\nn\\
&&\q\;2 A_{\rho\sigma\lambda\kappa} +2 A_{\rho\kappa\lambda\sigma}
\ea
Thus, $p_{\rho\sigma}, p_{\rho\sigma\kappa},p_{\rho\lambda\sigma\kappa}$ are expressed as linear functions of the area metric components $A_{\mu\nu\xi\zeta}$. There are (a priori) 6 independent quantities $p_{\rho\sigma}$, 12 independent quantities $p_{\rho\sigma\kappa}$, and  3 independent quantities $p_{\rho\lambda\sigma\kappa}$. But using their expression in terms of the area metric components we find the following constraint for the $p$-quantities 
\ba\label{3.5}
2\sum_{\rho<\sigma} p_{\rho\sigma}\q-\sum_{\sigma<\kappa,\sigma\neq\rho\neq\kappa} p_{\rho\sigma\kappa}\q + \sum_{0<\kappa<\sigma,0<\lambda} p_{0\lambda\sigma\kappa} &=& 0\q .
\ea
This constraint  is a reflection\footnote{The constraint represents a left null vector of the matrix describing the linear  equation (\ref{3.5}). Hence, there needs to be also a right null vector.} of the fact, that on the right hand side of (\ref{AM25}),  the area metric components appear only in the combinations $A_{0123}+A_{0321}$ and $A_{0213}+A_{0312}$, as well as combinations generated from these two by applying the symmetries (\ref{SymA}).  That is, the combination 
\ba
A_{0123}+A_{0231}+A_{0312} 
\ea
which appears in the cyclicity condition (\ref{Cycl3.3}) 
can, in particular, not be generated. We can thus not determine $A_{0123}+A_{0231}+A_{0312} $ by using the equations (\ref{AM25}).  The cyclicity condition sets this combination to zero, which allows us to determine the area metric from measuring areas only. 

In Area Regge calculus the area metric is reconstructed from the areas of parallelograms \cite{AR2}. It therefore leads to a cyclic area metric.

\section{The (modified) Plebanski action} \label{SecMPleb}

We will here provide a short review of the Plebanski action \cite{Plebanski}, in particular that it can be understood as a $BF$-action with simplicity constraints imposed. For a much more extensive review see e.g. \cite{KrasnovRev}. Following \cite{Krasnov1,FreidelMod,Spez,SpezB} we will then construct a modification of the theory, in which the constraints are replaced by potentials terms. 

The variables in the  (non-chiral) Plebanski action include an $\text{so}(4)$-valued two-form $B^{IJ}_{\mu\nu}$ and an $\text{SO}(4)$ connection $\omega^{IJ}_\mu$. The anti-symmetric index pair $(I,J)$ provides the labels for the $\text{so}(4)$ basis, with $I,J=0,\ldots,3$  and lower-case greek letters denote space-time indices $\mu=0,\ldots,3$.  We have furthermore  Lagrange multiplier fields $\phi_{IJKL}$, for which we impose the following symmetries $\phi_{IJKL}=-\phi_{JIKL}=-\phi_{IJLK}=\phi_{KLIJ}$ and $\phi_{IJKL}\epsilon^{IJKL}=0$.

The Plebanski action \cite{Plebanski} is then  given by
\ba\label{PA1}
S&=& \int     \delta_{IJKL}   B^{IJ} \wedge F^{KL}(\omega) \,+\, \frac{1}{2\gamma} \epsilon_{IJKL} B^{IJ} \wedge F^{KL}(\omega)\,-\,\tfrac{1}{2} \phi_{IJKL}B^{IJ}\wedge B^{KL} \q .
\ea
The first term defines a $BF$-action, the second term will lead to the so-called Holst term, which comes with the Barbero-Immirzi parameter $\gamma$. Without the third term, which imposes the simplicity constraints, this action is topological. The simplicity constraints, which are obtained by varying (\ref{PA1}) with respect to the Lagrange multipliers $\phi_{IJKL}$ (with the symmetries detailed above imposed), are given by
\ba\label{SC1}
B^{IJ} \wedge B^{KL}  = c\, \epsilon^{IJKL}
\ea
where $c$ is a four-form that can be determined by contracting (\ref{SC1}) with the Levi-Civita symbol, and is thus given by $c=\tfrac{1}{24} B^{IJ}\wedge B^{KL} \epsilon_{IJKL}$.

There are four sectors of solutions to the simplicity constraints, given by
\ba\label{SCSol}
B^{KL}= \pm  (\star e\wedge e)^{KL} \q ,\q\q   B^{KL}=\pm(e\wedge e)^{KL}
\ea
where the $e_\mu^I$ can be interpreted as  co-tetrads. Using the first set of solutions (with the $+$-sign) the Plebanski action does evaluate to the Palatini action with a Holst term \cite{Holst}.  The second set of solutions (\ref{SCSol}) defines the so-called topological sector. The reason for this name is that using these solutions in the first term in the action (\ref{PA1}) gives a topological action, which happens to be proportional to the Holst term.

The Plebanski action (\ref{PA1}) leads therefore to the Palatini (-Holst) action and thus to general relativity. Here one assumes that the simplicity constraints are imposed 'sharply'. Alternatively, one can replace the imposition of the simplicity constraints through Lagrange multipliers  by a `weaker' version, namely by suppressing the simplicity constraints with a potential.  To this end one replaces the  constraints terms in (\ref{PA1}) with \cite{Krasnov1,SmolinUni}
\ba\label{BPot}
\tfrac{1}{2} \phi_{IJKL}B^{IJ}\wedge B^{KL}  \q\rightarrow \tfrac{1}{2}\left( \phi_{IJKL}+\tfrac{1}{6} V(\phi) \epsilon_{IJKL}\right)B^{IJ}\wedge B^{KL}  \q .
\ea
Variation of this modified action with respect to the $\phi$-fields gives the equation of motion
\ba\label{EOMphi}
B^{IJ}\wedge B^{KL} \,=\, c\,\left( \epsilon^{IJKL}- 4 \frac{ \partial V(\phi)} {\partial \phi_{IJKL}}\right) \q ,
\ea
where $c=\frac{1}{24} B^{IJ}\wedge B^{KL}\epsilon_{IJKL}$. Assuming that the potential $V(\phi)$ admits a non-degenerate Hessian, one can solve (\ref{EOMphi}) for the $\phi$-fields in terms of the $B$-fields. Re-inserting these solutions into (\ref{BPot}) one obtains a potential term for the $B$-fields. If $V(\phi)$ has constant directions, one still obtains a corresponding subset of the simplicity constraints. The solutions to these subset of the simplicity constraints can be also re-inserted into (\ref{BPot}), leading to a potential term for the remaining parameters.

One can furthermore integrate out the connection from the (modified) Plebanski action, see \cite{Bengtsson, FreidelMod, Spez} for details. This leaves us with a second order action of the $B$-fields only, on which the full set or only a  subset of the simplicity constraints can be imposed. In the next Section \ref{Bparam}, we will introduce different parametrization of the (gauge invariant content of the) $B$-fields.  The first is in terms of a length metric and auxiliary fields, the second is in terms of area metrics. We can thus obtain actions in terms of these parametrizations.

\section{Parametrizations of $B$-fields}\label{Bparam}

\subsection{Parametrizations of  $B$-fields in terms of tetrads and auxiliary fields }

 The variables in the (non-chiral) Plebanski action
include the $\text{SO}(4)$ connection $\omega^{IJ}_\mu$. 
But this connection can be integrated out \cite{Bengtsson, FreidelMod, Spez},  and one can thus define a second order action in terms of the $B$-field only. The $\text{so}(4)$-algebra carries an (internal Hodge) dualization operator $(\star)^{IJ}_{KL}=\frac{1}{2}\epsilon^{IJ}_{KL}$, which satisfies $\star\star=\mathbb{I}$. The space of $\text{so}(4)$-valued two-forms can be split into $\pm$-eigenspaces with respect to the $\star$-operator, which are each three-dimensional. We will use the following parametrization for this split
\ba
B^{IJ}_{\mu \nu}= {P^{IJ}_{+}}_i (B^i_+)_{\mu\nu}+{P^{IJ}_{-}}_i (B^i_-)_{\mu\nu} \q, 
\ea
where 
\ba
{P^{IJ}_{\pm}}_i=\pm \delta^{IJ}_{0i}+\tfrac{1}{2} \epsilon^{IJ}_{0i}
\ea
and $i=1,2,3$.  We will use lower case latin letters for indices running from $1$ to $3$. Note that these can be used to label a basis in $\text{su}(2)$. Indeed the above split corresponds to the fact that $\text{so}(4)=\text{su}(2)_+ \oplus \text{su}(2)_-$.

We will now use a parametrization of the $B$-fields, which has been introduced in \cite{FreidelMod}.  We rewrite each of the $\text{su}(2)$-valued two-forms  $B^i_\pm$ in terms of a tetrad $(e_\pm)_\mu^I$ and eight (independent) scalars  that are organized into a $3\times 3$ matrix $(b_\pm)^i_j$ of determinant 1:
\ba\label{ParamB}
{B^i_\pm}_{\mu\nu}(b_\pm,e_\pm)= \sigma_\pm \, (b_\pm)^i_j  \,\, {\Sigma^j_\pm}_{\mu\nu}(e_\pm) \q .
\ea
Here $\sigma_\pm$ is a choice of sign, introduced for later convenience. The so-called Plebanski two-forms $\Sigma_\pm$ depend on a tetrad, and are defined as
\ba\label{defSig}
{\Sigma^j_\pm}_{\mu\nu}(e)&:=&\pm (e^0_\mu e^i_\nu-e^0_\nu e^i_\mu)+{\epsilon^i}_{jk}e^j_\mu e^k_\nu \,=\,2 {P^{IJ}_{\pm}}_i  \, \delta^j_i  \delta^{IJ}_{KL}  \,  e^K_\mu e^L_\nu  \q .
\ea
These two sets of three two-forms ${\Sigma^j_\pm}_{\mu\nu}(e)$ are self-dual, respectively antiself-dual, with respect to the space-time Hodge dual operator $(*)^{\mu\nu}_{\rho\sigma}=\frac{1}{2}g_{\rho\kappa}g_{\sigma\lambda} \epsilon^{\kappa\lambda\mu\nu}$, as determined by the (tetrad induced) length metric. In fact they define a basis for the three-dimensional space of (anti-) self-dual two-forms. 

We can recover the $\text{SO}(4)$ gauge invariant part of the parametrization (\ref{ParamB}) as follows \cite{FreidelMod}:
\begin{itemize}
\item
 The Urbantke metric \cite{Urbantke} does recover the conformal class of the space-time metric defined by the tetrad, that is
 \ba\label{Umetric}
 \sigma_\pm \det(e_\pm) {e_\pm}_\mu^I {e_\pm}_\nu^J\delta_{JK} &=& \pm  \frac{1}{12} \epsilon_{ijk} \epsilon^{\alpha \beta \kappa \lambda} {B_{\pm}}^i_{\mu\alpha} {B_\pm}^j_{\beta\kappa} {B_\pm}^k_{\kappa\nu}     \q .
 \ea
 \item
 The $(b_\pm)^i_j$ and the determinant $\det(e_\pm)$ are encoded in
 \ba\label{qmetric}
\det(e_\pm)  q^\pm_{ij}&:=& \det(e_\pm)\,\, (b_\pm)^k_i (b_\pm)_j^l \delta_{kl}\,=\,\pm \frac{1}{8} {B^i_\pm}_{\mu\nu} {B^j_\pm}_{\rho\sigma}\epsilon^{\mu\nu\rho\sigma}  \q .
 \ea
 Note that, because the $b$-matrices are uni-modular, we have  $\det(q^\pm)=1$.
\end{itemize}

The solutions 
$
B^{IJ}= \star (e\wedge e)^{IJ}     
$
to the simplicity constraints (\ref{SC1}) translate to the conditions
\ba
(e_+)^I_\mu=(e_-)^I_\mu=e^I_\mu, \q \q{b^i_+}_j={b^i_-}_j=\delta^i_j ,\q \q \sigma_\pm=\pm 1 \q .
\ea
That is, the two tetrads need to be identified with each other and the $b$-fields are frozen to the identity. The choice $\sigma_\pm=\pm 1$ defines the gravitational sector of the solution space to the simplicity constraints. One obtains the topological sector for $\sigma_+=\sigma_-$.

Let us count the components in the various parametrizations: We started out with $36$ components of the $B$ field, which split into $2\times 18$ components for the right- and left-handed parts respectively. Taking the $\text{SU}(2)$ gauge symmetry for each sector into account, we have $2\times 15$ invariant degrees of freedom. 

On the other hand, we have 16 tetrad components and 8 independent $b$-components in each sector. But the $\Sigma_\pm(e)$ do not change if we apply a  rotation in the subgroup $ \mathbb{I}\times \text{SU(2)}$, respectively  $\text{SU}(2)\times \mathbb{I}$. Furthermore, we have an additional invariance under $\text{SO}(3)$ rotations acting on the contracted index $j$ in the parametrization (\ref{ParamB}). We are thus left with 18 components in each sector.  Taking the remaining $\text{SU(2)}$ gauge symmetry into account, we find again  $2\times 15$ invariant degrees of freedom. 

\subsection{Area metrics from $B$-fields}

We now introduce for each sector an area metric given by \cite{Reisi}
\ba\label{Arpm}
A^\pm_{\mu\nu \rho\sigma} &=&\frac{1}{2} \left( {B^i_\pm}_{\mu\nu}   {B^j_\pm}_{\rho\sigma} \delta_{ij} - \tfrac{1}{4!} {B^i_\pm}_{\mu'\nu'}   {B^j_\pm}_{\rho'\sigma'} \delta_{ij} \epsilon^{\mu'\nu' \rho'\sigma'} \epsilon_{\mu\nu \rho\sigma}\right) \q .
\ea
The second term on the right hand side of (\ref{Arpm}) ensures that the area metrics are cyclic, that is, they satisfy $A^\pm_{\mu\nu \rho\sigma}\epsilon^{\mu\nu \rho\sigma}=0$.  

In terms of the parametrization (\ref{ParamB}) the area metrics are given by
\ba\label{Argq}
A^\pm_{\mu\nu \rho\sigma} &=&\frac{1}{2} \left( q^\pm_{ij}  {\Sigma^i_\pm}_{\mu\nu}\!(e_\pm)\,\,{\Sigma^j_\pm}_{\rho\sigma}\!(e_\pm) \,\,   \mp \tfrac{1}{3} \text{Tr}(q^\pm) \det(e_\pm)  \epsilon_{\mu\nu\rho\sigma}             \right)
\ea 
where $q^\pm_{ij}= (b_\pm)^k_i (b_\pm)_j^l \delta_{kl}$ are  symmetric matrices with determinant $\text{det}(q^\pm)=1$.

A cyclic area metric has 20 independent components. The area metrics are clearly  invariant under the $\text{SU}(2)$ action on the $B_\pm$-fields, but the $B_\pm$-fields contain each only 15 gauge invariant degrees of freedom.  Thus the area metrics $A^\pm$ do not provide a convenient parametrization for the gauge invariant content of the $B_\pm$-fields. However, it has been shown in \cite{Reisi}, that demanding 
\ba\label{LHRH}
A^+_{\mu\nu \rho\sigma} =A^-_{\mu\nu \rho\sigma}
\ea
is equivalent to the simplicity constraints, at least if one considers only the gravitational sector. Indeed (\ref{LHRH}) does give 20 independent ($\text{SO}(4)$ gauge invariant) equations, thus reducing the 30 gauge invariant $B$-field degrees of freedom to the 10  ($\text{SO}(4)$ gauge invariant) degrees of freedom of the length metric.

As we elaborated on in Section \ref{Intro}, we are looking for a subset of the simplicity constraints, so that their imposition leaves 20 ($\text{SO}(4)$ gauge invariant) degrees of freedom, that allow us to define an area metric. Clearly, demanding that the area metrics coincide, is too strong of a requirement. But the area metrics $A^\pm$ do encode the length metrics $g^\pm_{\mu\nu}=(e_\pm)^I_\mu(e_\pm)^J_\nu\delta_{IJ}$, defined by the tetrads $e_\pm$ appearing in the parametrization of the $B_\pm$-fields in (\ref{ParamB}). We can therefore demand that 
\ba
g^+_{\mu\nu}=g^-_{\mu\nu} \q , 
\ea
which gives 10 independent equations, and leaves 20 ($\text{SO}(4)$ gauge invariant) degrees of freedom. These can be organized into one area metric
\ba
A_{\mu\nu \rho\sigma}(e,q^+,q^-) =A^+_{\mu\nu \rho\sigma} (e,q^+)+A^-_{\mu\nu \rho\sigma} (e,q^-) \q ,
\ea
where $A^+$ and $A^-$ are defined with the same tetrad $e$, but with two sets of auxiliary fields $q^+$ and $q^-$. Note that the area metric is induced by the length metric  defined by the tetrad $e_\mu^I$, if one sets $q^\pm_{ij}=\delta_{ij}$:
\ba\label{ArInd}
A_{\mu\nu \rho\sigma}(e,\delta_{ij},\delta_{ij}) &=& 2  e_\mu^I e_\nu^J e_\rho^K e_\sigma^L \delta_{IJKL}\,=\,g_{\mu\rho}g_{\nu\sigma}-g_{\mu\sigma}g_{\nu\rho} \q .
\ea

 We now have a match between the 20 ($\text{SO}(4)$ gauge invariant) degrees of freedom given by the area metric, and the 20 ($\text{SO}(4)$ gauge invariant) degrees of freedom encoded in the $q^\pm$ fields and the tetrad $e$.

This allows (in principle) to obtain an action for area metrics, starting from the action for modified Plebanski theory:  To this end we integrate out the $\phi$-fields and the connection $\omega$ from the modified Plebanski action, and use the parametrization (\ref{ParamB}) for the $B$-fields. See \cite{Spez,SpezB} for the explicit expression for this action.   We then set the tetrads $e_+=e_-=e$ to be equal and obtain an action that can be written in terms of one metric field $g_{\mu\nu}=\delta_{IJ}e^I_\mu e^J_\nu$ and fields $q^\pm_{ij}$.  Inverting (\ref{Argq}) we can express $g_{\mu\nu}$ and $q^\pm_{ij}$ in terms of an area metric.

The inversion of (\ref{Argq}) requires  to solve a set 20 polynomial equations, and we cannot expect to find an explicit solution.\footnote{From this perspective, the existence of an explicit expression of the length metric and the $q^\pm$-field in terms of the $B$-fields, see (\ref{Umetric}) and (\ref{qmetric}) is rather surprising. However, in forming the area metric we loose access to the internal indices and to the split into $\pm$-sectors, making the inversion of (\ref{Argq}) much more difficult.}
The inversion can however be done perturbatively. Assuming a split $A={\bf A} + \varepsilon\, a$ of the area metric into background and perturbations, we expand $g={\bf g}+ \varepsilon^1 \, {}^{(1)}\!g+ \varepsilon^2 \, {}^{(2)}\! g+\cdots$ and $q^\pm = {\bf q}^\pm+\varepsilon^1\, {}^{(1)}\!q^\pm+ \varepsilon^2\, {}^{(2)}\! q+\cdots$  and can solve
\ba
{\bf A} + \varepsilon\, a &=& A(g,q^\pm)
\ea
for $g$ and $q$ order by order. Below we consider a linearization around flat space, leading to a quadratic action for the area metric fluctuations.

\subsection{Linearization}\label{SecLin}

We will consider a linearization around a configuration where the area metric is induced by the flat metric:
\ba
q^\pm_{ij} \,=\, \delta_{ij}+\chi_{ij}^\pm \;,\q\q   e^I_\mu \,=\, \delta^I_\mu + \eta^I_\mu \q , \q\q A_{\mu\nu\rho\sigma}=2\delta_{\mu[\rho}\delta_{\sigma] \nu} +a_{\mu\nu\rho\sigma}\q .
\ea
Note that $\det(q^\pm)=1$ implies to first order in the perturbations $\chi_{ij}\delta^{ij}=0$.  The tetrad fluctuations $\eta^I_\mu$ define the length metric fluctuations 
\ba
h_{\mu\nu}= \eta^I_\mu \delta_{I\nu}+\eta^I_\nu \delta_{I\mu}.
\ea
 We  now raise and lower indices with the flat space-time metric $\delta_{\mu\nu}$ and the internal metric $\delta_{ij}$.

The expansion of (\ref{Argq}) to first order in the perturbations leads to
\ba\label{ArExp}
a_{\mu\nu\rho\sigma}&=& \mathbb{L}^{\lambda\tau}_{\mu\nu\rho\sigma} h_{\lambda\tau} \,+\, 2\mathbb{P^+}^{ij}_{\mu\nu\rho\sigma} \chi^+_{ij}\,+\, 2\mathbb{P^-}^{ij}_{\mu\nu\rho\sigma} \chi^-_{ij}
\ea
where
\ba\label{defLP}
 \mathbb{L}^{\lambda\tau}_{\mu\nu\rho\sigma} &=& 
 2\delta^{~}_{\mu[\rho} \delta^{(\lambda}_{\sigma]} \delta^{\tau)}_{\nu^{\;}}
 -
  2\delta^{~}_{\nu[\rho} \delta^{(\lambda}_{\sigma]} \delta^{\tau)}_{\mu^{\;}}
 \nn\\
 \mathbb{P^\pm}^{ij}_{\mu\nu\rho\sigma} & =&  \tfrac{1}{2} \left( {P^\pm}^i_{\mu\nu}    {P^\pm}^j_{\rho\sigma}   + {P^\pm}^j_{\mu\nu}    {P^\pm}^i_{\rho\sigma}     \right) -\tfrac{1}{3} \delta^{ij}  {P^\pm}^{i'}_{\mu\nu}    {P^\pm}^{j'}_{\rho\sigma}\delta_{i'j'}  
\ea
with ${P^\pm}^i_{\mu\nu} =\tfrac{1}{2}\Sigma^i_{\pm \mu\nu}(
\delta_\rho^I)$. Here we implemented the fact that the $\chi_{ij}$ are symmetric and traceless into the definition of $\mathbb{P^\pm}$. The $\mathbb{P^\pm}$ tensors are furthermore (anti-) self-dual in both space-time index pairs with respect to the flat metric.

The part $\mathbb{L}^{\lambda\tau}_{\mu\nu\rho\sigma} h_{\lambda\tau}$ describes an area metric fluctuation induced by a length metric fluctuations and can be also obtained from expanding (\ref{ArInd}). The remaining parts in (\ref{ArExp}) describe area metric fluctuations that are orthogonal to the ones induced by the length metric. In fact, introducing the notation  $A \circ B = A_{\mu\nu\rho\sigma}  \delta^{\mu\mu'} \delta^{\nu\nu'} \delta^{\rho\rho'}\delta^{\sigma\sigma'} B_{\mu'\nu'\rho'\sigma'}$, we have for the contractions of the various tensors
\ba
\mathbb{L}^{\lambda\tau}  \circ \mathbb{L}^{\lambda'\tau'} &=& 8 \delta^{\lambda (\lambda'} \delta^{\tau') \tau}+ 4\delta^{\lambda\tau} \delta^{\lambda'\tau'}\,=:\, 8\,\mathbf{I}^{\lambda\tau\lambda'\tau'} + 4 \delta^{\lambda\tau} \delta^{\lambda'\tau'}=: K^{\lambda\tau \lambda\tau}\,\,, \nn\\
\mathbb{P^\pm}^{ij} \circ \mathbb{P^\pm}^{i'j'}&=& \delta^{i(i'} \delta^{j')j} -\tfrac{1}{3} \delta^{ij} \delta^{i'j'} \,=: {\mathbf I}^{ij i'j'}
\nn\\
\mathbb{L}^{\lambda\tau}  \circ \mathbb{P^\pm}^{ij} &=&0 , \,\, \nn\\
  \mathbb{P^+}^{ij} \circ \mathbb{P^-}^{kl}&=&0   \q .
\ea
The right hand side of the first equation can be understood as a Gram matrix $K$ acting on the space of symmetric rank two tensors. Its inverse is given by
\ba
(K^{-1})_{\lambda \tau \lambda'\tau'}&=&\tfrac{1}{8}\,\mathbf{I}_{\lambda\tau\lambda'\tau'} -\tfrac{1}{8\cdot 6} \delta^{\lambda\tau} \delta^{\lambda'\tau'}
\ea
The right hand side of the second equation does already define the identity on the space of symmetric, traceless rank-two tensors of dimension 3.

With these expressions at hand, we can invert the relation (\ref{ArExp}), and express the $h$ and $\chi^\pm$ fluctuations in terms of the area metric fluctuations
\ba\label{itrafo}
h_{\lambda \tau} &=& (K^{-1})_{\lambda \tau \lambda'\tau'}\,\mathbb{L}^{\lambda'\tau'} \circ a \,\, =:  \mathbb{K}_{\lambda\tau} \circ a   \,\, ,\nn\\
\chi^{\pm}_{ij} &=&   \tfrac{1}{2}    \mathbb{P^\pm}_{ij} \circ a       \q ,
\ea
with
\ba
((K^{-1})_{\lambda \tau \lambda'\tau'}\,\mathbb{L}^{\lambda'\tau'} )_{\mu\nu\rho\sigma}=\tfrac{1}{8}\mathbf{I}_{\lambda\tau\lambda'\tau'}\mathbb{L}^{\lambda'\tau'}_{\mu\nu\rho\sigma} -\tfrac{1}{8\cdot 3}\delta_{\lambda \tau} \mathbb{A}^{\rm S}_{\mu\nu\rho\sigma}     \q ,
\ea
where we defined $\mathbb{A}^{\rm S}_{\mu\nu\rho\sigma}\,=\,\delta_{\mu\rho}\delta_{\nu\sigma}-\delta_{\mu\sigma}\delta_{\nu\rho}$.

This also allows us to define projectors for the space of area metric fluctuations (see also Appendix \ref{AppA})
\ba
\Pi^{\rm L}_{\mu\nu\rho\sigma\,|\, \mu'\nu'\rho'\sigma'}\!\!&=&\!\mathbb{L}^{\lambda\tau }_{\mu\nu\rho\sigma} (K^{-1})_{\lambda \tau \lambda'\tau'}\,\mathbb{L}_{\mu'\nu'\rho'\sigma'}^{\lambda'\tau'} \nn\\
 &=& 
 \tfrac{1}{4\cdot 2} ( \mathbb{A}^+_{\mu\nu\mu'\nu'} \mathbb{A}^-_{\rho\sigma\rho'\sigma'} \!\!+ \mathbb{A}^+_{\mu\nu\rho'\sigma'} \mathbb{A}^-_{\rho\sigma\mu'\nu'}  )
 +\tfrac{1}{4\cdot 2} ( \mathbb{A}^-_{\mu\nu\mu'\nu'} \mathbb{A}^+_{\rho\sigma\rho'\sigma'} \!\!+ \mathbb{A}^-_{\mu\nu\rho'\sigma'} \mathbb{A}^+_{\rho\sigma\mu'\nu'}  )
 \! +\!\tfrac{1}{8\cdot 3} \mathbb{A}^{\rm S}_{\mu\nu\rho\sigma}\mathbb{A}^{\rm S}_{\mu'\nu'\rho'\sigma'}
\nn\\
\Pi^{\pm}_{\mu\nu\rho\sigma\,|\, \mu'\nu'\rho'\sigma'}\!\!&=& \!\mathbb{P^\pm}^{ij}_{\mu\nu\rho\sigma}  \mathbf{I}_{iji'j'} \mathbb{P^\pm}^{i'j'}_{\mu'\nu'\rho'\sigma'}  \nn\\
&=&
\tfrac{1}{4\cdot 2} ( \mathbb{A}^\pm_{\mu\nu\mu'\nu'} \mathbb{A}^\pm_{\rho\sigma\rho'\sigma'} \!\!+ \mathbb{A}^\pm_{\mu\nu\rho'\sigma'} \mathbb{A}^\pm_{\rho\sigma\mu'\nu'}  )\! -\!\tfrac{1}{4\cdot 3} \mathbb{A}^\pm_{\mu\nu\rho\sigma}\mathbb{A}^\pm_{\mu'\nu'\rho'\sigma'}\nn\\
\ea
where
\ba
\mathbb{A}^\pm_{\mu\nu\rho\sigma}&=&\tfrac{1}{2}{ P^\pm}^{i}_{\mu\nu} { P^\pm}^{j}_{\rho\sigma}\,=\, \tfrac{1}{2}(\delta_{\mu\rho}\delta_{\nu\sigma}-\delta_{\mu\sigma}\delta_{\nu\rho})\pm \tfrac{1}{2} \epsilon_{\mu\nu\rho\sigma} \q ,\q\q 
\mathbb{A}^{\rm S}_{\mu\nu\rho\sigma}=\mathbb{A}^+_{\mu\nu\rho\sigma}+\mathbb{A}^-_{\mu\nu\rho\sigma} \q .
\ea
These projectors sum to the identity on the space of cyclic area metrics: $\Pi^{\rm L}+\Pi^{+}+\Pi^{-}=\mathbf{I}$, see Appendix \ref{AppA}.

\section{Area metric actions from modified Plebanski actions}\label{SecAct}

In this section we will construct a linearized action for the area metric fluctuations. We will start with the linearization of the $BF$-part of the action  (\ref{PA1}), where the connection has been integrated out, and the parametrization (\ref{ParamB}) of the $B$-fields has been used. Here we will  impose that the tetrads of the $+$- and $-$-sector coincide $e_+=e_-=e$.

For notational convenience we will state a given mode contribution $S(k)$   in the Fourier transformed action 
\ba
S=\sum_k S(k)=\sum_k \phi(-k) H(k) \phi(k)\q ,
\ea
and use $\bar{\phi}(k)=\phi(-k)$ for the Fourier transformed fields.

The Fourier transform of the linearized $BF$-Lagrangian, with $h_+=h_-\equiv h$ set equal,  can be expressed as  \cite{FreidelMod,Spez}
\ba\label{Act1}
S_{1} (k)&=&  S^+_{BF}(k)  +  S^-_{BF}(k)\nn\\ 
&=&\tfrac{\gamma_+}{2}\,\, \left(\bar{h}_{\mu\nu} + \bar{\chi}^+_{\mu\nu} \right) {\cal E}^{\mu\nu\rho\sigma}  \left(h_{\rho\sigma}+ \chi^+_{\rho\sigma} \right) \,\,\, +\,\, \,  \tfrac{\gamma_-}{2} \,\, \left(\bar{h}_{\mu\nu} + \bar{\chi}^-_{\mu\nu} \right) {\cal E}^{\mu\nu\rho\sigma}  \left(h_{\rho\sigma}+ \chi^-_{\rho\sigma} \right) \q ,
\ea
where we introduced $\gamma_+=1+\tfrac{1}{\gamma}$ and $\gamma_-=1-\tfrac{1}{\gamma}$.
Here, ${\cal E}$ describes the linearized Einstein-Hilbert (or Fierz-Pauli) action. It can be expressed in terms of spin projectors as 
\ba\label{EHE}
 {\cal E}_{\mu\nu\rho\sigma}  &=& \tfrac{\Delta}{4} \left( {}^2\!P_{\mu\nu\rho\sigma}-2\,  {}^0\!P_{\mu\nu\rho\sigma}\right) \q ,
\ea
where $\Delta=k_\mu k_\nu \delta^{\mu\nu}$ is the (positive definite) Laplacian and ${}^2\!P$ and ${}^0\!P$ project onto the symmetric, transverse, traceless  tensor modes and symmetric, transverse, trace  tensor modes respectively. See Appendix \ref{AppB} for their explicit definition. 

Furthermore, we have  introduced in (\ref{Act1}) an isometric embedding of the $\chi^\pm_{ij}$ into the space of  symmetric, transverse, traceless tensors, given by
\ba\label{chitensor}
\chi^\pm_{\mu \nu} =   E_{\mu\nu}^{\pm ij}\chi^\pm_{ij} \,:=\, 4 \, {\mathbb P}^{\pm ij}_{\mu\rho \nu\sigma}  \frac{k^\rho k^\sigma} {\Delta}  \chi^\pm_{ij} \q 
\ea
with ${\mathbb P}^{\pm ij}_{\mu\rho \nu\sigma}$ defined in (\ref{defLP}). 
The isometricity condition means that
$
\bar{\chi}^\pm_{\mu \nu} {\bf I}^{\mu\nu\mu'\nu'}{\chi}^\pm_{\mu' \nu'}\,=\, \bar{\chi}^\pm_{ij} \, {\bf I}^{iji'j'} \, \chi^\pm_{i'j'}
$
where ${\bf I}^{\mu\nu\rho\sigma}=\tfrac{1}{2} (\delta^{\mu\rho}\delta^{\nu\sigma}+ \delta^{\mu\sigma}\delta^{\nu\rho})$ is the projector onto symmetric tensors $t_{\mu\nu}$ and ${\bf I}^{iji'j'}=\tfrac{1}{2}(\delta^{ii'}\delta^{jj'}+\delta^{ij'}\delta^{ji'}) -\tfrac{1}{3}\delta^{ij}\delta^{i'j'}$ is the projector onto symmetric, traceless tensors $s_{ij}$.

Thus the ${}^2\!P$-projector in ${\cal E}$ acts as an identity on the $\chi^\pm_{\mu\nu}$ and ${}^0\!P$ annihilates the $\chi^\pm_{\mu\nu}$. We have therefore 
\ba\label{Act2}
S_{BF}^\pm(k) &=& \tfrac{1}{2}\gamma_\pm \,\,\left(\,\,
\bar{h}_{\mu\nu} {\cal E}^{\mu\nu\rho\sigma}  h_{\rho\sigma} 
\,+\, \tfrac{ \Delta}{4} \bar{h}_{\mu\nu}  (\chi^\pm)^{\mu\nu}+\tfrac{ \Delta}{4} \bar{\chi}^\pm_{\mu\nu} h^{\mu\nu} \,+\, \tfrac{ \Delta}{4} \bar{\chi}^\pm_{\mu\nu} (\chi^\pm)^{\mu\nu}
\,\,\right)
\ea

The version (\ref{Act1}) of the chiral actions $S_{BF}^\pm(k)$ makes it obvious, that the (separate) actions are invariant under the following shift symmetry parametrized by  traceless, symmetric tensors $\zeta^\pm_{ij}$
\ba
h_{\mu\nu} \rightarrow h_{\mu\nu} -   E_{\mu\nu}^{\pm ij} \zeta^\pm_{ij} , \q\q \chi^\pm_{ij}  \rightarrow \chi^\pm_{ij}  +\zeta^\pm_{ij} \q .
\ea
In fact this symmetry follows from the shift symmetry of the $BF$-action, which renders it to be topological.  In addition, we have the usual (linearized) diffeomorphism symmetry, given by adding longitudinal modes to the linearized metric field
\ba
h_{\mu\nu} \rightarrow  h_{\mu\nu} + k_\mu v_\nu + k_\nu v_\mu \q .
\ea
Accordingly, we have $9=5+4$ independent gauge symmetries for  $S_{BF}^+(k)$  and $9=5+4$ independent gauge symmetries for $S_{BF}^-(k)$. 

However, adding $S_{BF}^+$ and $S_{BF}^-$ (and identifying the linearized metrics in both sectors), the two shift symmetries combine to one symmetry, which can be parametrized by
\ba
h_{\mu\nu} \rightarrow h_{\mu\nu} -   E_{\mu\nu}^{+ ij} \zeta_{ij} , \q\q \chi^+_{ij}  \rightarrow \chi^+_{ij}  +\zeta_{ij} \, ,\q\q  
\chi^-_{ij}  \rightarrow \chi^-_{ij} + {\bf I}_{ijmn}\, E^{-mn}_{\rho\sigma}\, {\bf I}^{\rho\sigma\mu\nu}\, E^{+kl}_{\mu\nu} \zeta_{kl} 
\q .
\ea
Note that 
\ba
E^{-ij}_{\lambda\tau}{\bf I}_{ijmn}\, E^{-mn}_{\rho\sigma}\, {\bf I}^{\rho\sigma\mu\nu}\, E^{+kl}_{\mu\nu} =E^{+ij}_{\lambda\tau}{\bf I}_{ijmn}\,,
\ea
 as $E^{-ij}{\bf I}_{ijmn}\, E^{-mn}$ equates to the ${}^2\!P$-projector, which acts as identity on the $E^{+kl}$-tensors.
 
 We therefore have also $9=5+4$ independent gauge symmetries for $S_1$.
 
 The 5 shift symmetries will be broken by adding mass terms to the action. Assuming an $\text{SO}(4)$ invariant potential for the modified Plebanski action, it can -- at quadratic order -- only lead to mass terms involving  the norms of the $\chi^\pm$ fields \cite{SpezB}, that is
 \ba
 S_m(k)=S^+_m(k)+S^-_m(k)
 \,=\, \tfrac{ \gamma_+m_+^2}{8}   \bar{\chi}^+_{\mu\nu} (\chi^+)^{\mu\nu} \,+\,  \tfrac{\gamma_-m_-^2 }{8}  \bar{\chi}^-_{\mu\nu} (\chi^-)^{\mu\nu} 
 \,=\, \tfrac{\gamma_+m_+^2}{8}  \, \bar{\chi}^+_{ij} \,{\bf I}^{iji'j'} \chi^+_{i'j'} \,+\,  \tfrac{\gamma_-m_-^2 }{8} \, \bar{\chi}^-_{ij}  \,{\bf I}^{iji'j'}\chi^-_{i'j'}  \, .\,\,\;\;\;
 \ea
  These mass terms suppress the $\chi^\pm$ fluctuations, which parametrize the deviation of the  area metric from one that is induced by a length metric. They can therefore be understood as one form of weak imposition of the (remaining) simplicity constraints.

 We define the modified Plebanski action for the area metric fluctuations as 
$
 S_{\rm ModP}=S_{1}+S_m
 $.
 Using the variable transformations (\ref{itrafo}) this action can be expressed in matrix notation as 
\ba
S_{\rm ModP}(k)&=& 
 \bar{a}  \bigg(\! 
  \tfrac{ \gamma_+ \Delta }{8}  
  \left( ({}^2\!P-2{}^0\!P) \mathbb{K}+\mathbb{E}^+        \right)^t \left( ({}^2\!P-2{}^0\!P) \mathbb{K}+\mathbb{E}^+        \right)+
    \tfrac{ \gamma_- \Delta }{8}  
  \left(( {}^2\!P -2{}^0\!P) \mathbb{K}+\mathbb{E}^-       \right)^t \left(( {}^2\!P -2{}^0\!P) \mathbb{K}+\mathbb{E}^-       \right) +
\nn\\
&&\q\q\tfrac{\gamma_+m_+^2}{8\cdot 4 }  \Pi^++ \tfrac{\gamma_-m_-^2}{8\cdot 4 }  \Pi^-
\! \bigg)a \nn\\
 &=&\bar{a}  \left(\! 
  {\mathbb K}^t \, {\cal E} \,  {\mathbb K}       + 
 \tfrac{ \gamma_+ \Delta }{8} 
 \left( \! (\mathbb{E}^+ )^t {\mathbb K} + {\mathbb K}^t \mathbb{E}^+  + \tfrac{1}{2}  \Pi^+\! \right)
+ \tfrac{ \gamma_-\Delta }{8} 
\left(\! (\mathbb{E}^- )^t {\mathbb K} + {\mathbb K}^t \mathbb{E}^-  + \tfrac{1}{2}  \Pi^- \!\right)
+\tfrac{\gamma_+m_+^2}{8\cdot 4 }  \Pi^++ \tfrac{\gamma_-m_-^2}{8\cdot 4 }  \Pi^-
\!\right) a    \, , \nn\\
\ea
where
\ba
\mathbb{E}^{\pm}_{\lambda\tau \,|\,\mu\nu\rho\sigma}&=&E^{\pm ij}_{\lambda \tau} {\mathbf I}_{iji'j'} \mathbb{P}^{\pm i' j'}_{\mu\nu\rho\sigma} 
\,=\,4\,\, \Pi^{\pm}_{\lambda\kappa\tau \eta \,|\,\mu\nu\rho\sigma}\frac{k^\kappa k^\eta}{\Delta}
\,\, ,   
\ea
and $\Pi^\pm ={\mathbb P}^{\pm ij}{\mathbf I}_{ijkl} {\mathbb P}^{\pm kl}$ are projectors on the space of tensors of rank 4.
Note that 
\ba
\mathbb{E}^{\pm}_{\lambda\tau \,|\, \mu\nu\rho\sigma} {\mathbf I}^{\lambda \tau \kappa \eta}\,  h_{\kappa \eta} &=& 2 \,\,\Pi^\pm_{\mu\nu\rho\sigma  \,|\,\kappa \eta \lambda \tau}\, {}^{(1)}\! R^{\kappa \eta \lambda \tau}\,=:\, 2\,\,  {}^{(1)}\! W^\pm_{\mu\nu\rho\sigma}
\ea
where 
\ba
{}^{(1)}\!R^{\kappa \eta \lambda \tau}&=& k^\kappa k^{[\lambda} h^{\tau] \eta} - k^\eta k^{[\lambda} h^{\tau]\kappa}
\ea
is the first order perturbation of the Riemann tensor. Thus ${}^{(1)}\! W^\pm$ are the first order perturbations of the self-dual and anti-self dual part of the Weyl tensor.  That is, the $BF$-actions couple the $\chi^\pm$-fields to the  self-dual and antiself-dual part of the Weyl tensor, respectively.

The explicit expression of $S_{\rm ModP}(k)$ using products and contractions of Kronecker-deltas, Levi-Civita-symbols and momenta $k_\mu$ is very lengthy, due to the many algebraic symmetries that have to be implemented. But we provide in Appendix \ref{AppC} an expression using projections of tensors onto the space of symmetric bi-linear maps for area metric perturbations.

\subsection{The effective action for the length metric}

Integrating out the $\chi^\pm$ variables we obtain an effective action for the length variables. The solution for the $\chi^\pm$ are given by
\ba
\chi^\pm_{ij}&=&-\frac{ \Delta}{m^2_\pm + \Delta} \, {\bf I}_{ijkl} \, E^{\pm kl}_{\mu \nu}\,  {\bf I}^{\mu\nu \rho \sigma}\,  h_{\rho \sigma} \q .
\ea
Inserting these solutions into the Lagrangian $L_{\rm ModP}$ we use 
\ba\label{EId}
E^{\pm ij}_{\mu\nu}{\bf I}_{ijkl} E^{\pm kl}_{\rho\sigma}={}^2\!P_{\mu\nu\rho\sigma}
\ea
 and  obtain
\ba\label{effL}
S_{\rm eff}(k)&=& \bar{h}_{\mu\nu} {\cal E}^{\mu\nu\rho\sigma}  h_{\rho\sigma}  - \frac{1}{8}\left(  \frac{\gamma_+}{m^2_+ + \Delta} +\frac{\gamma_-}{m^2_-+ \Delta}\right) \Delta^2\,\, \bar{h}_{\mu\nu} \,{}^2\!P^{\mu\nu\rho\sigma} h_{\rho \sigma}    \q .
\ea

 Note that 
\ba
\Delta^2\,\, \bar{h}_{\mu\nu} \,{}^2\!P^{\mu\nu\rho\sigma} h_{\rho \sigma} \,=\, 2\,\, {}^{(1)}\! \bar{W}_{\mu\nu\rho\sigma} \,\, {}^{(1)}\! W^{\mu\nu\rho\sigma}
\ea
where $ {}^{(1)}\! W= {}^{(1)}\! W^++ {}^{(1)}\! W^-$ is the first order perturbation of the Weyl curvature. We have thus a correction to the Einstein-Hilbert term, given by the square of the Weyl-tensor.  This correction term is suppressed for small energies  as compared to the mass squares $m^2_\pm$. The correction term is non-local, but this non-locality effects only terms of  sixth and higher order in a derivative expansion.

Let us remark that we obtain the couplings $\gamma_\pm$ for the square of the self-dual part of the Weyl tensor and the square of the antiself-dual part of the Weyl tensor --- the sum of these squares coincides with the square of the full Weyl tensor. The mechanism behind  is that the square of the embedding maps $E^{\pm ij}$, as  defined in (\ref{EId}),  gives the spin-2 projector for both the $+$- and the $-$-sector.

Note also, that we get the same kind of effective action, if we would consider only the self-dual part or the anti-self-dual part of the $BF$-action with mass terms for the $\chi$-fields added, see also \cite{Krasnov2}.  (In this case one does not have sufficient fields to obtain an area metric with 20 independent components though.)  As discussed in depth in \cite{FreidelMod,Krasnov2} this special case allows a field redefinition $(h_{\mu\nu},\chi^\pm_{ij})\rightarrow (H_{\mu\nu}=h_{\mu\nu}+ \chi^{\pm}_{\mu\nu},\chi^\pm_{ij})$, so that only the (redefined) graviton is propagating --- the $\chi^\pm_{ij}$-field only appears with a mass term and without a coupling to $H_{\mu\nu}$, and can thus be integrated out, leaving us with the (linearized) Einstein-Hilbert action in the $H$-fields. This field redefinition is however non-local due to the embeddings  $\chi^{\pm}_{\mu\nu}=E^{\pm ij}_{\mu\nu}\chi^{\pm}_{ij}$. 

In any case, for energies $k^2\ll m^2_\pm$ the Einstein-Hilbert term will dominate.  As we explained in the introduction, the masses parametrize the suppression of non-metric degrees of freedom, which appear in spin foams due to an anomaly in the simplicity constraint algebra. One way to understand the reason for this anomaly \cite{EffSF1} is due to the discreteness of the area spectra, with the area gap given by the Planck area. It is therefore reasonable to assume that the masses $m_\pm$ are of the order of the Planck mass.

\section{An alternative area metric action}\label{SecAR}

An alternative way to derive a (continuum) area metric action is from the continuum limit of the linearized Area Regge action. It has been shown in \cite{AR2}, that this limit is described by a Lagrangian of the following form\footnote{In \cite{AR2} only 18 of the 20 independent components of the area metric are reconstructed from the lattice variables - the remaining variables are integrated out. But the geometric interpretation \cite{AR2} of the Area Regge action supports that the Hessian is of the form (\ref{AR1}).}
\ba\label{AR1}
S_{\rm AR}(k)&=& \bar{a}_{\mu\nu\rho \sigma} \, H_{\rm AR}^{\mu\nu\rho \sigma \,|\,\lambda\kappa\tau\eta} \, a_{\lambda\kappa\tau\eta}
\ea
where
\ba\label{HAR1}
  H^{\mu\nu\rho \sigma \,| \alpha \beta}&:=&    H_{\rm AR}^{\mu\nu\rho \sigma \,|\,\lambda\kappa\tau\eta} \, \mathbb{L}_{\lambda\kappa\tau\eta}^{\alpha \beta} \,\,=\,\, - \frac{1}{16}  \cdot \frac{1}{8} \left(\epsilon^{\mu\nu \alpha \tau }\epsilon^{\rho\sigma \beta\eta} +\epsilon^{\mu\nu \beta \tau }\epsilon^{\rho\sigma \alpha \eta}\right) k_\tau k_\eta \,+\,{\cal O}(k^3)  \q ,\nn\\
 H^{iji'j'}_\pm&:=&     4\,\, \mathbb{P^\pm}_{\mu\nu\rho \sigma }^{ij}  H_{\rm AR}^{\mu\nu\rho \sigma \,|\,\lambda\kappa\tau\eta} \, \mathbb{P^\pm}_{\lambda\kappa\tau\eta}^{i'j'} \,\,\approx\,\,  M^2 \,  {\bf I}^{iji'j'} \,+\,{\cal O}(k^2)
\ea
with $M$ a mass scaling with the square of the inverse lattice constant. In the second line we expressed the fact that $H^{iji'j'}$ does not reproduce the identity exactly, which is conjectured \cite{AR2} to be due to lattice artifact effects.  In the following we will assume that $H^{iji'j'}= M^2 {\bf I}^{iji'j'} +{\cal O}(k^2)$.

Here we adopt the same type of variable transformation as in (\ref{ArExp}), that is 
\ba\label{ArExp2}
a_{\mu\nu\rho\sigma}&=& \mathbb{L}^{\lambda\tau}_{\mu\nu\rho\sigma} h_{\lambda\tau} \,+\, 2\mathbb{P^+}^{ij}_{\mu\nu\rho\sigma} \chi^+_{ij}\,+\, 2\mathbb{P^-}^{ij}_{\mu\nu\rho\sigma} \chi^-_{ij} \q .
\ea

Note that the first line in (\ref{HAR1}) means that the Area Regge action reproduces ($-1/8$ times)  the dualized linearized Riemann tensor
\ba
  H^{\mu\nu\rho \sigma \,|\, \alpha \beta} h_{\alpha \beta} &=&- \frac{1}{8}  \cdot \frac{1}{8} ( \epsilon^{\mu\nu \lambda \kappa}\epsilon^{\rho\sigma \tau \eta} +\epsilon^{\mu\nu \tau \eta}\epsilon^{\rho\sigma \lambda \kappa}) \,\,{}^{(1)} R_{\lambda\kappa \tau \eta}  \q .
\ea
We furthermore have 
\ba
H^{\xi\zeta  \, \alpha \beta}&:=& \mathbb{L}^{\xi\zeta}_{\mu\nu\rho \sigma} H_{\rm AR}^{\mu\nu\rho \sigma \,|\,\lambda\kappa\tau\eta} \, \mathbb{L}_{\lambda\kappa\tau\eta}^{\alpha \beta}\, \,\,\;\;\,\,\,=\, {\cal E}^{\xi\zeta \alpha \beta} \nn\\
H_\pm^{ ij \, \alpha \beta}&:=&2\,\, {\mathbb{P}^{\pm}}^{ij}_{\mu\nu\rho \sigma} H_{\rm AR}^{\mu\nu\rho \sigma \,|\,\lambda\kappa\tau\eta} \, \mathbb{L}_{\lambda\kappa\tau\eta}^{\alpha \beta}\,=- \frac{1}{8} \Delta E^{\pm ij}_{\mu\nu} \mathbf{I}^{\mu\nu \alpha \beta}      \q .
\ea

The expressions (\ref{HAR1}) leave the terms of order $k^2$ in $H_\pm^{ijkl}$ unspecified.  We now determine these terms  by adopting a similar structure for $L_{\rm AR}$ as for $L_{\rm ModP}$. That is, the Lagrangian splits into a $+$- and a $-$-part, which can each be written as a complete square plus an additional mass term for the $\chi^\pm$ variables. With this assumption, we obtain
\ba
S_{\rm AR}(k)&=& \,\, \;\;\tfrac{1}{2} \,\left(\,\,
\bar{h}_{\mu\nu} {\cal E}^{\mu\nu\rho\sigma}  h_{\rho\sigma} 
\,-\, \tfrac{ \Delta}{4} \bar{h}_{\mu\nu}  (\chi^+)^{\mu\nu}     -\tfrac{ \Delta}{4} \bar{\chi}^+_{\mu\nu} h^{\mu\nu} 
\,+\, \tfrac{ \Delta}{4} \bar{\chi}^+_{\mu\nu} (\chi^+)^{\mu\nu} 
\,\,\right) \,+\, M^2 \bar{\chi}^+_{\mu\nu} (\chi^+)^{\mu\nu} \nn\\
&&+\tfrac{1}{2} \,\left(\,\,
\bar{h}_{\mu\nu} {\cal E}^{\mu\nu\rho\sigma}  h_{\rho\sigma} 
\,-\, \tfrac{ \Delta}{4} \bar{h}_{\mu\nu}  (\chi^-)^{\mu\nu}-\tfrac{ \Delta}{4} \bar{\chi}^-_{\mu\nu} h^{\mu\nu} \,+\, \tfrac{ \Delta}{4} \bar{\chi}^-_{\mu\nu} (\chi^-)^{\mu\nu} 
\,\,\right) \,+\, M^2 \bar{\chi}^-_{\mu\nu} (\chi^-)^{\mu\nu} \nn\\
&=&
\,\, \;\;\tfrac{1}{2}\,\, \left(\bar{h}_{\mu\nu} - \bar{\chi}^+_{\mu\nu} \right) {\cal E}^{\mu\nu\rho\sigma}  \left(h_{\rho\sigma}- \chi^+_{\rho\sigma} \right) 
\,+\, M^2 \bar{\chi}^+_{\mu\nu} (\chi^+)^{\mu\nu} \nn\\
 &&+  \tfrac{1}{2} \,\, \left(\bar{h}_{\mu\nu} - \bar{\chi}^-_{\mu\nu} \right) {\cal E}^{\mu\nu\rho\sigma}  \left(h_{\rho\sigma}- \chi^-_{\rho\sigma} \right)\,+\, M^2 \bar{\chi}^-_{\mu\nu} (\chi^-)^{\mu\nu} \q ,
\ea
where we used the mapping (\ref{chitensor}) of the $\chi^\pm_{ij}$ into the space of transversal, traceless symmetric tensor modes.

The reader can find an expression for $S_{\rm AR}(k)$ in terms of (symmetrizations of) products of Kronecker-deltas, Levi-Civita-symbols and momenta $k_\mu$ in Appendix \ref{AppC}.

We see that $L_{\rm AR}$ differs from $L_{\rm ModP}$  (with $\gamma_+=\gamma_-=1$, that is in the limit $\gamma \rightarrow \infty$, and with $m^2_\pm=8M^2)$ by a variable transformation $\chi_\pm^{ij} \rightarrow -\chi_\pm^{ij}$ for the parts of the area metric that are not induced by the length metric fluctuations. Integrating out the $\chi^\pm$ variables one will find the same effective Lagrangian (\ref{effL}) for both cases.

 Although one could therefore see the two actions as equivalent, the $\chi^\pm$ variables do have the same geometric meaning for both actions. That is, in case one could access the expectation values\footnote{As we consider a theory with diffeomorphism symmetry, one would have to construct gauge invariant observables which encode information on the $\chi\pm$ observables. This is in principle possible, see  e.g.\cite{PartObs} and also \cite{PartObs3} for a perturbative framework to construct such observables. Note that the $\chi^\pm$ are already invariant under linearized diffeomorphisms.} of the $\chi^\pm$ by measurements, one would be able to detect the difference in these actions. 

This difference is puzzling but can be seen as an indication that there is, apart from the modified Plebanski action, a second family of actions for area metrics, which leads to the same effective action for the length metric, and thus reproduces general relativity in the limit of large mass parameters.

\section{Discussion}\label{Disc}

Spin foams are path integrals based on the Plebanski formulation.  A key mechanism in this formulation is the imposition of simplicity constraints onto the geometric fields, so that these reduce to the length metric. The quantization procedure renders part of the simplicity constraint algebra anomalous, thus forcing a weak imposition of this part of the constraints. One should thus understand spin foams as a path integral over an enlarged configuration space. 

Our central conjecture is here, that this enlarged configuration space can be interpreted as the space of area metrics.  This conjecture is inspired by recent  results \cite{AR2} on the continuum limit of Area Regge calculus, which plays a crucial role for the understanding of the continuum limit of effective spin foams. 

 We therefore hope that one can define effective actions that capture the continuum limit of spin foams directly from the Plebanski action. But as spin foams do only impose part of the simplicity constraints, we constructed  a  modified Plebanski action in which a part of the simplicity constraints are relaxed such that an area metric remains as free variable. This allows us to define area metric actions, whose dynamical content we analyzed using a linearization. 

The area metric variables admit a natural split into a length metric and additional variables. We can thus integrate out the additional variables, and obtain in this way an effective action for the length metric. This effective action coincides (up to terms of order 4 in derivatives) with the one found in \cite{AR2}, confirming our approach.   At the level of the area metrics, the modified Plebanski action and the action found in \cite{AR2} almost coincide --- they differ in the sign for the coupling between the length metric and the additional variables.  Thus, the (linearized) actions can be mapped to each other by a variable transformation, which flips the sign for the additional variables.

As long as we cannot probe directly the additional variables in the area metric (e.g. via a matter coupling that would see these additional variables, this sign difference has no physical consequences. It is nevertheless puzzling, as the additional variables have the same geometrical interpretation for both actions.  

It will be interesting to see how other spin foam related actions, e.g. the Area-Angle action \cite{AreaAngle} fit in. Here one should also relax the part of the simplicity constraints described by the shape matching constraints that reduce the geometric configurations to the ones of Length Regge calculus. We believe that an analysis of the continuum limit of the resulting action, along the lines of \cite{AR1,AR2}, will lead to a similarly structured action. 

We have here defined an action for cyclic area metrics, which have 20 independent components. To this end we imposed a set of 10 independent simplicity constraints sharply.   Another possibility is to impose a slightly smaller set of 9 simplicity constraints sharply, leaving us with 21 independent components for an acyclic area metric. Here one could choose to require that only the conformal classes of the two length metrics, appearing in the parametrization of the $B$-fields (\ref{ParamB}) agree. We leave this for future research.

One assumption underlying this work is that the continuum limit of spin foams can be defined and leads to a theory related to the Plebanski framework, which spin foams set out to quantize. There are of course many open questions to show this at the non-perturbative level. This includes ensuring the convergence for the highly oscillatory path integral (see e.g. \cite{ADP} for progress), possibly fine-tuning parameters in the spin foam amplitudes to reach a sensible phase or a phase transition (see e.g. \cite{ContLimit} for methods and progress), the question of how to regain diffeomorphism symmetry, which is broken by the regularization used to define the spin foam path integral (see e.g. \cite{Improved}), or how to coarse grain the fields \cite{CG}.

But with the assumption that this can be achieved, we do expect that the enlargement of the configuration space from length to area metrics forced by the anomalous constraint algebra leads to physical effects, that may allow to distinguish spin foams from other purely length metric based ones.\footnote{As argued in \cite{DittrichRyan3}, one consequence of this enlargement is that different values of the Barbero-Immirzi parameter define unitarily inequivalent observable algebras.}

In this work we found that the enlargement of the configuration space to area metrics leads to a correction for the Einstein-Hilbert action. This correction is given by (minus) the Weyl curvature squared term. It is however crucial, that this term is suppressed in a non-local way, that is, by the inverses of $(M^2+\Delta)$ terms. If we would just have the mass term $M^2$ we would obtain a ghost (i.e. a massive pole with negative residue in the propagator), whereas the non-local form (\ref{effL}) of the action avoids\footnote{ A quadratic action with Hessian  $k^2-k^4/(M^2+k^2)$ leads to a propagator $1/k^2   + 1/M^2$.} such  ghosts.

This non-local nature of our effective length action (\ref{effL}) only appears at order six in the derivative expansion --- up to terms of order four the action (\ref{effL}) coincides with Einstein-Weyl gravity.  There is a range of literature on the phenomenology of Einstein-Weyl gravity, which might also apply to our case.
 E.g.  the absence of no-hair theorems  \cite{NoHair} in Einstein-Weyl gravity has recently inspired an extensive search for numerical solutions. For instance, \cite{Lu} studied spherically symmetric and asymptotically flat black holes in vacuum Einstein-Weyl gravity, showing that there exist classes of solutions which will generally differ from the Schwarzschild geometry. The inclusion of a thin-shell matter source was considered in \cite{Holdom}, whereas \cite{Bonanno1} focused on connecting asymptotic solutions with series expansions around the horizon, which were extended to wormhole geometries in \cite{Bonanno2}.

These examples illustrate that there may be a rich phenomenology associated to the  enlargement of the configuration space from length to area metrics in spin foams, and more specifically to the appearance of the Weyl squared term in the effective action for the length metric. We therefore hope that our work opens up new avenues for the exploration of the phenomenology of spin foams.

\appendix

\section{ Projectors on the space of area metric perturbations}\label{AppA}

Here we summarize the construction of the projectors $\Pi^L$ and  $\Pi^\pm$ from Section \ref{SecLin}.

The $\Pi^\pm$ are constructed out of the following maps
\ba
{P^\pm}^i_{\mu\nu}&=&\pm \delta^0_{[\mu}\delta_{\nu]}^{i}+\tfrac{1}{2} \epsilon_{0i\mu\nu} \q ,
\ea
which are combined to
\ba
 \mathbb{P^\pm}^{ij}_{\mu\nu\rho\sigma} & =&  \tfrac{1}{2} \left( {P^\pm}^i_{\mu\nu}    {P^\pm}^j_{\rho\sigma}   + {P^\pm}^j_{\mu\nu}    {P^\pm}^i_{\rho\sigma}     \right) -\tfrac{1}{3} \delta^{ij}  {P^\pm}^{i'}_{\mu\nu}    {P^\pm}^{j'}_{\rho\sigma}\delta_{i'j'}   \q .
\ea
The $\Pi^\pm$ are then defined as
\ba\label{App3}
\Pi^{\pm}_{\mu\nu\rho\sigma\,|\, \mu'\nu'\rho'\sigma'}
&=& \!\mathbb{P^\pm}^{ij}_{\mu\nu\rho\sigma}  \mathbf{I}_{iji'j'} \mathbb{P^\pm}^{i'j'}_{\mu'\nu'\rho'\sigma'}  \nn\\
&=&
\tfrac{1}{4\cdot 2} ( \mathbb{A}^\pm_{\mu\nu\mu'\nu'} \mathbb{A}^\pm_{\rho\sigma\rho'\sigma'} \!\!+ \mathbb{A}^\pm_{\mu\nu\rho'\sigma'} \mathbb{A}^\pm_{\rho\sigma\mu'\nu'}  )\! -\!\tfrac{1}{4\cdot 3} \mathbb{A}^\pm_{\mu\nu\rho\sigma}\mathbb{A}^\pm_{\mu'\nu'\rho'\sigma'} \q .
\ea
To verify the second line in Equation ({\ref{App3}) it is helpful to use
\ba\label{App4}
{P^\pm}^i_{\mu\nu} {P^\pm}_{\rho\sigma}^j\delta_{ij}&=& 2\mathbb{A}^\pm_{\mu\nu\rho\sigma} \,=\,  \tfrac{1}{4} ( \delta_{\mu\rho}\delta_{\nu\sigma}-\delta_{\mu\sigma}\delta_{\nu\rho} \pm \epsilon_{\mu\nu\rho\sigma}) \q .
\ea
The identity on the space of cyclic area metrics can be expressed as
\ba\label{App5}
\mathbf{I}_{\mu\nu\rho\sigma\,|\, \mu'\nu'\rho'\sigma'} &=&
\tfrac{1}{4\cdot 2}( \mathbb{A}^{\rm S}_{\mu\nu\mu'\nu'} \mathbb{A}^{\rm S}_{\rho\sigma\rho'\sigma'} \!\!+ \mathbb{A}^{\rm S}_{\mu\nu\rho'\sigma'} \mathbb{A}^{\rm S}_{\rho\sigma\mu'\nu'}  ) - \tfrac{1}{8\cdot 3} \mathbb{A}^{\rm D}_{\mu\nu\rho\sigma}\mathbb{A}^{\rm D}_{\mu'\nu'\rho'\sigma'}
\ea
where
\ba
\mathbb{A}^{\rm S}_{\mu\nu\rho\sigma}=\mathbb{A}^+_{\mu\nu\rho\sigma}+\mathbb{A}^-_{\mu\nu\rho\sigma} \,=\,(\delta_{\mu\rho}\delta_{\nu\sigma}-\delta_{\mu\sigma}\delta_{\nu\rho}) \q,\q\q
\mathbb{A}^{\rm D}_{\mu\nu\rho\sigma}=\mathbb{A}^+_{\mu\nu\rho\sigma}-\mathbb{A}^-_{\mu\nu\rho\sigma} \,=\, \epsilon_{\mu\nu\rho\sigma}  \q .
\ea
Thus the first term on the right hand side of (\ref{App5}) does implement the symmetries (\ref{SymA}) of the area metrics and the second term  implements the projection onto the space of cyclic area metrics.

We can now define a projector $\Pi^L$, which is orthogonal to $\Pi^+$ and $\Pi^-$ by
\ba
\Pi^L _{\mu\nu\rho\sigma\,|\, \mu'\nu'\rho'\sigma'} &=&(\mathbf{I} -\Pi^+ -\Pi^-)_{\mu\nu\rho\sigma\,|\, \mu'\nu'\rho'\sigma'} \nn\\
&=&
 \tfrac{1}{4\cdot 2} ( \mathbb{A}^+_{\mu\nu\mu'\nu'} \mathbb{A}^-_{\rho\sigma\rho'\sigma'} \!\!+ \mathbb{A}^+_{\mu\nu\rho'\sigma'} \mathbb{A}^-_{\rho\sigma\mu'\nu'}  )
 +\tfrac{1}{4\cdot 2} ( \mathbb{A}^-_{\mu\nu\mu'\nu'} \mathbb{A}^+_{\rho\sigma\rho'\sigma'} \!\!+ \mathbb{A}^-_{\mu\nu\rho'\sigma'} \mathbb{A}^+_{\rho\sigma\mu'\nu'}  )
 \! +\!\tfrac{1}{8\cdot 3} \mathbb{A}^{\rm S}_{\mu\nu\rho\sigma}\mathbb{A}^{\rm S}_{\mu'\nu'\rho'\sigma'} \, .\nn 
\ea
This projector can furthermore be split into a trace part   and  a trace-less part
\ba
\Pi^{L,tr}&=&\tfrac{1}{8\cdot 3} \mathbb{A}^{\rm S}_{\mu\nu\rho\sigma}\mathbb{A}^{\rm S}_{\mu'\nu'\rho'\sigma'} \nn\\
\Pi^{L,ntr}&=&\tfrac{1}{4\cdot 2} ( \mathbb{A}^+_{\mu\nu\mu'\nu'} \mathbb{A}^-_{\rho\sigma\rho'\sigma'} \!\!+ \mathbb{A}^+_{\mu\nu\rho'\sigma'} \mathbb{A}^-_{\rho\sigma\mu'\nu'}  )
 +\tfrac{1}{4\cdot 2} ( \mathbb{A}^-_{\mu\nu\mu'\nu'} \mathbb{A}^+_{\rho\sigma\rho'\sigma'} \!\!+ \mathbb{A}^-_{\mu\nu\rho'\sigma'} \mathbb{A}^+_{\rho\sigma\mu'\nu'}  ) \q .
\ea

\section{ Spin projectors for length metric perturbations}\label{AppB}

Here we introduce the spin projectors for symmetric tensors of rank 2 and dimension $d$. They are given by
\ba\label{grav2}
{}^0\!P_{\mu\nu\rho\sigma}  &=&  \frac{1}{d-1} P^{\perp}_{\mu\nu} P^{\perp}_{\rho\sigma}  \nn\\
{}^1\!P_{\mu\nu\rho\sigma} &=& \frac{1}{2}(\delta_{\mu\rho}\delta_{\nu\sigma}+\delta_{\mu\sigma}\delta_{\nu\rho}) \,-\,\frac{1}{2}(P^{\perp}_{\mu\rho}P^{\perp}_{\rho\sigma}+P^{\perp}_{\mu\sigma}P^{\perp}_{\nu\rho})  \nn\\
{}^2\!P_{\mu\nu\rho\sigma}&=& \frac{1}{2}(P^{\perp}_{\mu\rho}P^{\perp}_{\rho\sigma}+P^{\perp}_{\mu\sigma}P^{\perp}_{\nu\rho}) -  \frac{1}{d-1} P^{\perp}_{\mu\nu} P^{\perp}_{\rho\sigma} 
\ea
where
\ba
P^{\perp}_{\mu\nu} =\delta_{\mu\nu}-\frac{k_\mu k_\nu}{\Delta}  \q,\q\q \Delta=k_\mu k_\nu \delta^{\mu\nu}
\ea
is the projector onto the transversal vectors. As can be easily seen ${}^0\!P,{}^1\!P$ and ${}^2\!P$ sum to the identity map (on the space of symmetric rank two tensors)
and are orthogonal to each other.

\section{Area metric Hessian tensors}\label{AppC}

The algebraic symmetries of the area metrics lead to lengthy expressions for the area metric action, if expressed with contraction of Kronecker Delta's and Levi-Civita tensors. More compact expressions can be obtained, if we leave the symmetrizations implicit.

We project a general tensor $t^{\alpha\beta\gamma\delta\,|\,\mu\nu\rho\sigma}$ to its part  $T^{\alpha\beta\gamma\delta \,|\,\mu\nu\rho\sigma}$, which is
\begin{itemize}
\item anti-symmetric in the index pairs $(\alpha\beta),(\gamma\delta), (\mu\nu),(\rho\sigma)$,
\item symmetric under exchange of index pairs $(\alpha\beta)\leftrightarrow(\gamma\delta)$ and $(\mu\nu)\leftrightarrow(\rho\sigma)$ and
\item
symmetric under exchange of the index quadruples $(\alpha\beta\gamma\delta)\leftrightarrow(\mu\nu\rho\delta)$,
\end{itemize}
by defining
\ba
T^{\alpha\beta\gamma\delta\,|\,\mu\nu\rho\sigma} &:=&  \frac{1}{2^7}\Bigg\{ \Bigg\{\Big\{ \big\{ t^{\alpha\beta\gamma\delta\,|\,\mu\nu\rho\sigma} -[\mu\leftrightarrow\nu] -  [\rho\leftrightarrow\sigma] + [\mu\leftrightarrow\nu, \rho\leftrightarrow\sigma]\big\}-[\alpha \leftrightarrow \beta] - [\gamma\leftrightarrow\delta] + [\alpha\leftrightarrow\beta, \gamma\leftrightarrow\delta] \Big\}\nn\\
 &{}&{\;\;\;\;\;\;\;\;\;}+ [\alpha\beta\leftrightarrow\gamma\delta]+[\mu\nu\leftrightarrow\rho\sigma]+[\alpha\beta\leftrightarrow\gamma\delta,\,\mu\nu\leftrightarrow\rho\sigma ]\Bigg\}
+ [\alpha\beta\gamma\delta \leftrightarrow \mu\nu\rho\sigma]\Bigg\}\,.
 \ea
Finally, we project $T^{\alpha\beta\gamma\delta\,|\,\mu\nu\rho\sigma}$ onto the subset of tensors cyclic in the first four and last four indices via
\ba
\hat{T}^{\alpha\beta\gamma\delta\,|\,\mu\nu\rho\sigma} &:=& T^{\alpha\beta\gamma\delta\,|\,\mu\nu\rho\sigma} - \frac{1}{4!} \epsilon^{\alpha\beta\gamma\delta}\epsilon_{\alpha'\beta'\gamma'\delta'}T^{\alpha'\beta'\gamma'\delta'\,|\,\mu\nu\rho\sigma} -  \frac{1}{4!}T^{\alpha\beta\gamma\delta\,|\,\mu'\nu'\rho'\sigma'}\epsilon_{\mu'\nu'\rho'\sigma'}\epsilon^{\mu\nu\rho\sigma}\nn\\
&{}&+\frac{1}{4!}\frac{1}{4!}\epsilon^{\alpha\beta\gamma\delta}\epsilon_{\alpha'\beta'\gamma'\delta'}T^{\alpha'\beta'\gamma'\delta'\,|\,\mu'\nu'\rho'\sigma'}\epsilon_{\mu'\nu'\rho'\sigma'}\epsilon^{\mu\nu\rho\sigma},
\ea 

In this way we construct tensors $\hat T_A$ with $A=0,l\dots,7$ from tensors $t_A$, given by
\ba
t_0^{\alpha\beta\gamma\delta\,|\,\mu\nu\rho\sigma}  &=& \delta^{\alpha\mu}\delta^{\beta\nu}\delta^{\gamma\rho}\delta^{\delta\sigma}, \\
t_1^{\alpha\beta\gamma\delta\,|\,\mu\nu\rho\sigma}  &=& \delta^{\alpha\gamma} \delta^{\mu\rho}\delta^{\beta\nu}\delta^{\delta\sigma}, \\
t_2^{\alpha\beta\gamma\delta\,|\,\mu\nu\rho\sigma}  &=& \delta^{\alpha\gamma} \delta^{\beta\delta}\delta^{\mu\rho}\delta^{\nu\sigma},\\
{t}_3^{\alpha\beta\gamma\delta\,|\,\mu\nu\rho\sigma}  &=& \epsilon^{\alpha\beta\mu\nu} \delta^{\gamma\rho}\delta^{\delta\sigma} ,\\
\ea
and
\ba
t_4^{\alpha\beta\gamma\delta\,|\,\mu\nu\rho\sigma}  &=& \delta^{\alpha\mu} \delta^{\gamma\rho} \delta^{\nu\sigma} k^\beta k^\delta,\\
{}\q\q\q\; t_5^{\alpha\beta\gamma\delta\,|\,\mu\nu\rho\sigma}  &=& \delta^{\alpha\gamma}\delta^{\mu\rho}\delta^{\nu\sigma} k^\beta k^\delta,\\
t_6^{\alpha\beta\gamma\delta\,|\,\mu\nu\rho\sigma}  &=& \delta^{\alpha\gamma} \delta^{\beta\nu} \delta^{\mu\rho} k^\delta k^\sigma,\\
t_{7}^{\alpha\beta\gamma\delta\,|\,\mu\nu\rho\sigma}  &=& {\epsilon^{\alpha\beta\mu}}_{\lambda}\delta^{\gamma\rho}\delta^{\nu\sigma}k^{\lambda}k^\delta \q .
\ea

The previous definitions allow us to express the Hessian tensors for the linearized area metric actions defined by $S(k) = \bar{a}_{\alpha\beta\gamma\delta} H^{\alpha\beta\gamma\delta\,|\,\mu\nu\rho\sigma}a_{\mu\nu\rho\sigma}$ in a compact form.
For the area metric actions derived from the modified non-chiral Plebanski action and from the Area-Regge action the Hessians respectively take the form
\ba
H_{\rm ModP} &=&
 \frac{\Delta}{32}\hat{T}_0 - \frac{\Delta}{8}\hat{T}_1 + \frac{\Delta}{48}\hat{T}_2  +
 \frac{1}{4}\hat{T}_4 -\frac{1}{12}\hat{T}_5+\frac{1}{8}\hat{T}_6 + 
\frac{\gamma_+-\gamma_-}{2}\left(\frac{\Delta}{64}\hat{T}_3 + \frac{1}{8}\hat{T}_{7}\right) \nn\\ 
&&+\q \frac{\gamma_+m^2_++\gamma_-m^2_-}{32}\left(  \frac{1}{2} \hat T_0- \hat T_1 +\frac{1}{6} T_2 \right)
+\frac{\gamma_+m^2_+-\gamma_-m^2_-}{ 32\cdot 4} \hat T_3\,\;
, \\ 
H_{\rm AR } &=&
 \frac{\Delta}{32}\hat{T}_0 + \frac{\Delta}{8}\hat{T}_1 - \frac{\Delta}{16}\hat{T}_2 
 -\frac{1}{4}\hat{T}_4+ \frac{1}{4}\hat{T}_5  - \frac{3}{8}\hat{T}_6 + 
\frac{\gamma_+-\gamma_-}{2}\left(\frac{\Delta}{64}\hat{T}_3 - \frac{1}{8}\hat{T}_{7}\right)   \nn\\
&&
+\q \frac{\gamma_+m^2_++\gamma_-m^2_-}{32}\left( \frac{1}{2} \hat T_0- \hat T_1 +\frac{1}{6} T_2\right)
+\frac{\gamma_+m^2_+-\gamma_-m^2_-}{ 32\cdot 4} \hat T_3 \q .
\ea
Here we generalized $H_{\rm AR}$ by introducing different couplings for the self-dual and antiself-dual sector, in a similar way as for $H_{\rm ModP}$. To match the couplings from Section \ref{SecAR}, we have to choose $\gamma^{-1}=0$ (hence $\gamma_+=\gamma_-=1$) and $\tfrac{m^2_+}{8}=\tfrac{m_-^2}{8}=M^2$.

\begin{acknowledgments}
  JNB thanks Benjamin Knorr for discussions on higher-derivative gravity. BD thanks  Hal Haggard, Ted Jacobson and  Wolfgang Wieland for interesting discussions.  BD and JNB thank in particular José Padua-Argüelles for enlightening communication.   JNB is supported by  NSERC grants awarded to BD and Niayesh Afshordi.
Research at Perimeter Institute is supported in part by the Government of Canada through the Department of Innovation, Science and Economic Development Canada and by the Province of Ontario through the Ministry of Colleges and Universities.
\end{acknowledgments}

\vspace{1cm}

\begingroup
\endgroup

\end{document}